\documentclass[twocolumn]{aastex631}

\usepackage{xspace}
\usepackage{hyperref}
\usepackage{multirow}
\usepackage{graphicx}
\usepackage{enumitem}
\usepackage{amssymb}
\usepackage{amsmath}
\usepackage{float}
\usepackage{array}

\newcommand{\chandra}{\textit{Chandra}\xspace}
\newcommand{\xmm}{XMM-\textit{Newton}\xspace}
\defcitealias{lamassa16}{L16}

\shorttitle{Stripe 82-XL}
\shortauthors{Peca et al.}

\begin{document}

\title{Stripe 82-XL: the $\sim$54.8 deg$^2$ and $\sim$18.8 Ms \chandra and \xmm point source catalog and number of counts}

\email{$^{\star}$peca.alessandro@gmail.com}

\author[0000-0003-2196-3298]{Alessandro Peca$^{\star}$}
\affiliation{Department of Physics, University of Miami, 
Coral Gables, FL 33124, USA}

\author[0000-0002-1697-186X]{Nico Cappelluti}
\affiliation{Department of Physics, University of Miami, 
Coral Gables, FL 33124, USA}

\author[0000-0002-5907-3330]{Stephanie LaMassa}
\affiliation{Space Telescope Science Institute, 3700 San Martin Drive, Baltimore, MD 21210, USA}

\author[0000-0002-0745-9792]{C. Megan Urry}
\affiliation{Yale Center for Astronomy \& Astrophysics, 52 Hillhouse Avenue, New Haven, CT 06511, USA}
\affiliation{Department of Physics, Yale University, P.O. Box 208120, New Haven, CT 06520, USA}

\author[0000-0001-7220-0755]{Massimo Moscetti}
\affiliation{Vanderbilt University, Nashville, TN 37240, USA}
\affiliation{Department of Physics, University of Miami, 
Coral Gables, FL 33124, USA}

\author[0000-0001-5544-0749]{Stefano Marchesi}
\affiliation{Dipartimento di Fisica e Astronomia (DIFA), Università di Bologna, via Gobetti 93/2, I-40129 Bologna, Italy}
\affiliation{Department of Physics and Astronomy, Clemson University, Kinard Lab of Physics, Clemson, SC 29634-0978, USA}
\affiliation{INAF - Osservatorio di Astrofisica e Scienza dello Spazio di Bologna, Via Piero Gobetti, 93/3, 40129, Bologna, Italy}

\author[0000-0002-1233-9998]{David Sanders}
\affiliation{Institute for Astronomy, University of Hawaii, 2680 Woodlawn Drive, Honolulu, HI 96822, USA}

\author[0000-0002-5504-8752]{Connor Auge}
\affiliation{Institute for Astronomy, University of Hawaii, 2680 Woodlawn Drive, Honolulu, HI 96822, USA}

\author[0000-0002-2525-9647]{Aritra Ghosh}
\affiliation{DiRAC Institute and the Department of Astronomy, University of Washington, Seattle, WA 98195, USA}

\author[0000-0000-0000-0000]{Tonima Tasnim Ananna}
\affiliation{Department of Physics and Astronomy, Wayne State University, Detroit, MI 48202, USA}

\author[0000-0003-3638-8943]{N\'uria Torres-Alb\`a}
\affiliation{Department of Physics and Astronomy, Clemson University, Kinard Lab of Physics, Clemson, SC 29634-0978, USA}

\author[0000-0007-7568-6412]{Ezequiel Treister}
\affiliation{Instituto de Astrof\'isica, Facultad de F\'isica, Pontificia Universidad Cat\'olica de Chile, Casilla 306, Santiago 22, Chile}

\begin{abstract}
We present an enhanced version of the publicly-available Stripe 82X catalog (S82-XL), featuring a comprehensive set of 22,737 unique X-ray point sources identified with a significance $\gtrsim 4\sigma$. This catalog is four times larger than the original Stripe 82X catalog, by including additional archival data from the \chandra and \xmm telescopes. Now covering $\sim54.8$ deg$^2$ of non-overlapping sky area, the S82-XL catalog roughly doubles the area and depth of the original catalog, with limiting fluxes (half-area fluxes) of 3.4$\times 10^{-16}$ (2.4$\times 10^{-15}$), 2.9$\times 10^{-15}$ (1.5$\times 10^{-14}$), and 1.4$\times 10^{-15}$ (9.5$\times 10^{-15}$) erg s$^{-1}$ cm$^{-2}$ across the soft (0.5-2 keV), hard (2-10 keV), and full (0.5-10 keV) bands, respectively. 
S82-XL occupies a unique region of flux-area parameter space compared to other X-ray surveys, identifying sources with rest-frame luminosities from $1.2\times 10^{38}$ to $1.6\times 10^{47}$ erg s$^{-1}$ in the 2-10 keV band (median X-ray luminosity, $7.2\times 10^{43}$ erg s$^{-1}$), and spectroscopic redshifts up to $z\sim6$. By using hardness ratios, we derived Active Galactic Nuclei (AGNs) obscuration obtaining a median value of $N_H=21.6_{-1.6}^{+1.0}$, and an overall, obscured fraction ($\log N_H/\mathrm{cm^{-2}}>22$) of $\sim 36.9\%$.  S82-XL serves as a benchmark in X-ray surveys and, with its extensive multiwavelength data, is especially valuable for comprehensive studies of luminous AGNs.
\end{abstract}


\section{Introduction} \label{sec:intro}
The exploration of Active Galactic Nuclei (AGNs, \citealp{antonucci93,urry95}) through multi-wavelength observations has significantly advanced our understanding of their intricate physical mechanisms. X-ray emission, in particular, serves as a crucial diagnostic tool for AGNs, offering insights into the phenomena surrounding the accretion of matter by supermassive black holes (SMBHs). The penetrating nature of X-ray photons allows them to reveal the innermost regions of AGNs, even in environments with high column densities of obscuring material \citep[e.g.,][]{hickox18}.
Over the past decades, X-ray surveys have become instrumental in unraveling the evolutionary pathways of AGNs across cosmic time. These surveys vary in scope and scale, from deep, small-area observations that unveil faint, distant objects (e.g., Chandra deep fields,  \citealp{xue16,luo17,liu17}, and COSMOS, \citealp{hasinger07,cappelluti07,cappelluti09, civano16,marchesi16a}), to extended large-area surveys important for detecting rare, luminous AGNs. Ideally, surveys would encompass wide areas at faint flux limits. Yet, current technological constraints compel astronomers to navigate a trade-off between the two. In particular, the scarcity of high-luminosity AGNs requires extensive surveys to sample large volumes of the Universe. Such coverage is critical for gathering statistically-robust samples that enable in-depth analyses and studies (e.g., \citealp{ananna19,peca23,auge23}). 
Among the notable large-area X-ray surveys ($\gtrsim10$ deg$^2$) are, for example, the \xmm $\sim7.1$ deg$^2$ of the \textit{Herschel}-ATLAS field (\citealp{ranalli15}); the Chandra Deep Wide-field Survey (CDWFS, \citealp{masini20}) covering $\sim$9.3 deg$^2$ of \chandra observations; the \xmm XXL North and South surveys \citep{pierre16,chiappetti18} which include two fields of $\sim$ 25 deg$^2$; Stripe 82X (S82X, hereafter) which covers $\sim 31.3$ deg$^2$ of \xmm and \chandra observations \citep{lamassa13a,lamassa13b,lamassa16}; eFEDS over $\sim 140$ deg$^2$ with eROSITA \citep{brunner22}; ExSeSS \citep{delaney23}, with an area of $\sim2086.6$ deg$^2$ observed by the \textit{Swift} telescope; and the recent first release of the eROSITA half-sky survey (eRASS:1, \citealp{merloni24}). These large surveys are pivotal in constructing a representative census of rare, high-luminosity AGN populations, facilitating a deeper understanding of their distribution, properties, and evolutionary paths.

S82X covers the Stripe 82 Legacy field (\citealp{frieman08}), a 300 deg$^2$ equatorial region scanned multiple times by the Sloan Digital Sky Survey (SDSS). The strength of Stripe 82 lies in its extensive multi-wavelength coverage, which spans a large portion of the entire electromagnetic spectrum, including: the UV (GALEX; \citealp{morrissey07}), optical (SDSS coadded catalogs; \citealp{jiang14,fliri16}), near-infrared (UKIDSS and VHS,  \citealp{hewett06,lawrence07}), mid-infrared  (WISE, \citealp{wright10}), and radio (FIRST, \citealp{helfand15}). 
Stripe 82 was also observed extensively in the near/mid/far-infrared with facilities no longer available, such as \textit{Spitzer} and \textit{Herschel} (\citealp{viero14,timlin16,papovich16}). Finally, X-ray observations continue to be added to the Stripe 82 archive, with \chandra (\citealp{lamassa13a}) and \xmm (\citealp{lamassa13b,lamassa16}). Thus, Stripe 82 provides an invaluable legacy of astronomical data (see \citealt{lamassa16}, hereafter \citetalias{lamassa16}, and \citealp{ananna17} for an in-depth overview, and Peca et al. in prep. for updates on multi-wavelength coverage). 
Given the heterogeneous nature of the X-ray coverage, it is crucial to keep this dataset current with newly available observations. 

This work presents a new X-ray catalog in the Stripe 82 field, ``Stripe 82-XL" (S82-XL), a significant enhancement to the existing S82X catalog, achieved by incorporating newly available archival observations from the \chandra and \xmm telescopes to the S82X dataset. By adding new observations, S82-XL improves the depth and expands the covered area of the earlier S82X catalog. This enhancement allows for a deeper exploration across AGN luminosity, redshift, and absorption, marking a substantial advance in X-ray extragalactic surveys.

The paper is organized as follows. In \S \ref{sec:data} we introduce the X-ray datasets used in this work. We describe the source selection in \S \ref{sec:selection}. S82-XL source fluxes and sky number densities are presented in \S \ref{sec:catalog}, while in \S \ref{sec:properties} we show the observed properties of the sources. In \S \ref{sec:summary} we summarize the work and discuss its implications in the field of extragalactic surveys. The catalog is described in \S \ref{app:cat_description}.
Throughout this paper, we assumed a $\Lambda$CDM cosmology with the fiducial parameters $H_0=70$ km s$^{-1}$ Mpc$^{-1}$, $\Omega_m=0.3$, and $\Omega_{\Lambda}=0.7$.

\section{Data}\label{sec:data}
To incorporate the most recent X-ray observations into the new S82-XL catalog, we used two state-of-the-art catalogs that extensively cover the Stripe 82 area: the second release of the Chandra Source Catalog (CSC, \citealp{evans10}) and the 4XMM Source Catalog (\citealp{webb20,traulsen20}). 

The CSC catalog aggregates data from various observations, systematically stacking overlapping fields to provide the highest possible sensitivity to detected sources. For the purposes of this study, we utilized the CSC version 2.1\footnote{\href{https://cxc.cfa.harvard.edu/csc/}{https://cxc.cfa.harvard.edu/csc/}.}, noting that while it is not yet complete at the moment of writing, the pending observations pertain primarily to the Galactic center and thus do not impact our analysis of the Stripe 82 region\footnote{CXC help desk, private communication.}. We queried the CSC on ``master level" sources, which are unique X-ray sources obtained from the combination of detections across multiple observations and stacks. The CSC v2.1 encompasses observations up until December 31, 2021, across multiple energy bands, namely: ultra-soft (u, 0.2$-$0.5 keV), soft (s, 0.5$-$1 keV), medium (m, 1$-$2 keV), hard (h, 2$-$7 keV), broad (b, 0.5$-$7 keV) for the ACIS-I and ACIS-S cameras, and wide (w, 0.1$-$10 keV) for the high-resolution camera (HRC).

The 4XMM catalog is available for individual (namely, 4XMM; \citealp{webb20}) and stacked observations (4XMM stacked, or 4XMMs; \citealp{traulsen20}), with the latter encompassing sources detected in overlapping fields. For this work, we used the latest iteration of the catalog at the moment of writing, 4XMM-DR12\footnote{\href{http://xmmssc.irap.omp.eu/Catalogue/4XMM-DR12/4XMM_DR12.html}{http://xmmssc.irap.omp.eu/Catalogue/4XMM-DR12/4XMM\_DR12.html}.}, along with the associated stacked catalog\footnote{\href{https://xmmssc.aip.de/cms/catalogues/4xmm-dr12s/}{https://xmmssc.aip.de/cms/catalogues/4xmm-dr12s/}.} to ensure a thorough inclusion of X-ray sources. The 4XMM catalog delineates sources across a broad spectrum of energy bands, namely: band 1 (0.2$-$0.5 keV), 2 (0.5$-$1 keV), 3 (1$-$2 keV), 4 (2$-$4.5 keV), 5 (4.5$-$12 keV), 8 (0.2$-$12 keV), and 9 (0.5$-$4.5 keV), thus complementing the CSC's dataset and enhancing the source selection. Like the CSC v2.1, 4XMM-DR12 includes observations up until December 31, 2021.

\section{Point source selection}\label{sec:selection}
We queried the CSC v2.1 and 4XMM-DR12 catalogs in the area $300\leq$ RA $\leq60$ and $-1.5\leq$ Dec $\leq+1.5$ degrees. 
This is slightly wider in declination than the nominal Stripe 82 ($-1.25<$ Dec $<+1.25$) in order to include areas with available X-ray sources and to exploit the available multiband coverage footprints (Peca et al. in prep.). Inside this region, there are currently 374 \xmm and 335 \chandra archival observations, for a total exposure of $\sim$18.79 Ms ($\sim$13.23 and $\sim$5.56 Ms for \xmm and \chandra, respectively). The total, non-overlapping area covered is $\sim54.8$ deg$^2$, with $\sim44.5$ and $\sim15.6$ deg$^2$ covered individually by \xmm and \chandra, respectively---roughly doubling the original S82X catalog \citep{lamassa13a,lamassa13b,lamassa16}.

To select point sources, we used a similar approach for both the CSC and 4XMM. First, we excluded extended sources as defined in both catalogs. For CSC sources, this was determined using the \textsc{extent\_flag} set to \textsc{False} at the master source level, indicating the source does not significantly deviate from a point source profile at a 90\% confidence level across any science energy band or contributing observation.  
Similarly, for 4XMM we selected source with \textsc{EP\_EXTENT}=0 (\textsc{EXTENT}=0 for 4XMMs), meaning that the source extension is less than 6\arcsec\, in the EP, ``observation-level", 0.2-12.0 keV band. This parameter aggregates measurements from the PN, MOS1, and MOS2 cameras, averaging them to assess the combined source extension.

\begin{figure}[!tp]
    \centering \vspace{0.5cm}
    \includegraphics[scale=0.45]{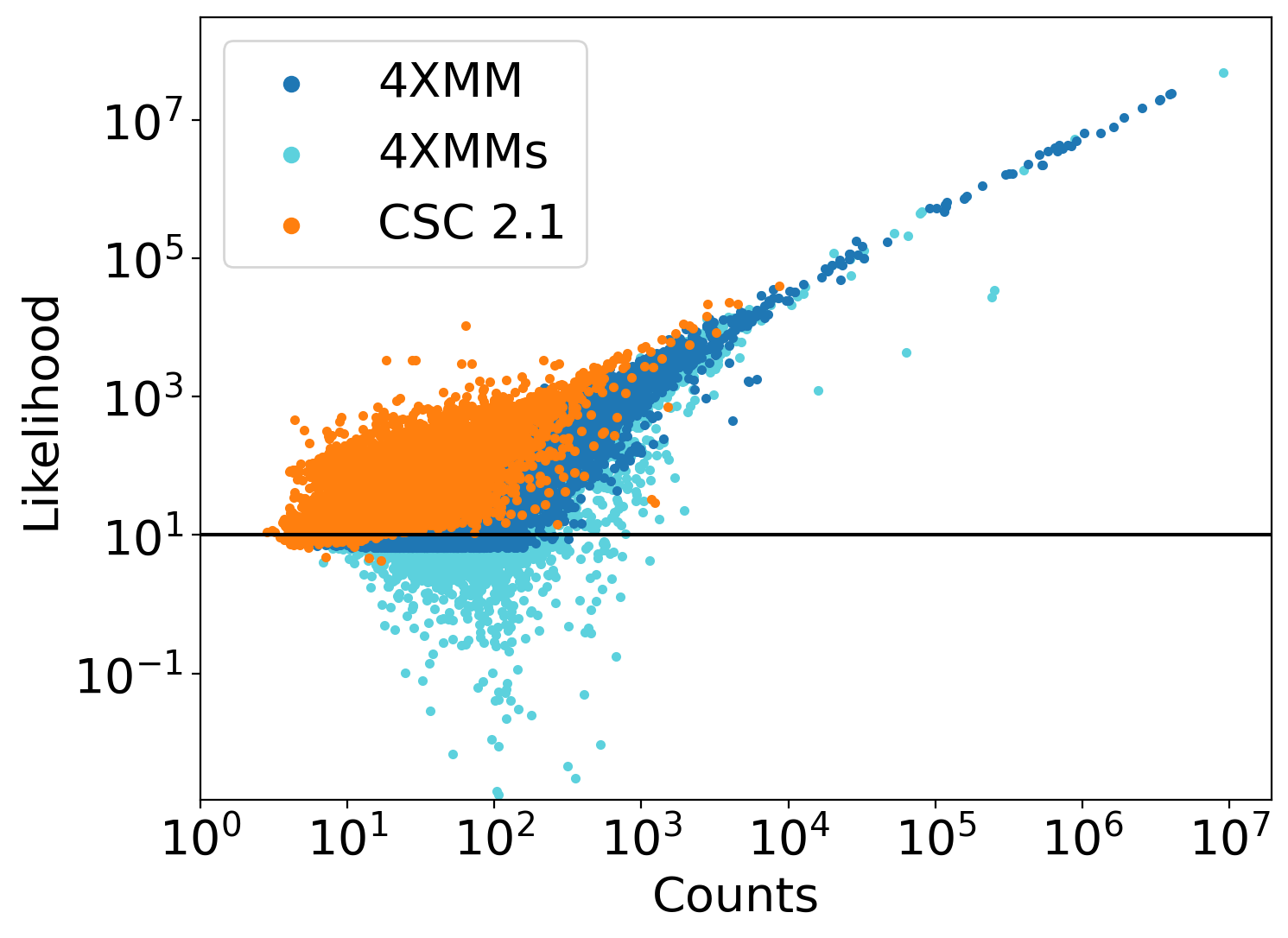}
    \caption{Detection likelihoods for 4XMM point sources (\textsc{EP\_8\_DET\_ML}; standard catalog in {\it blue} and stacked catalog in {\it cyan}), and for CSC point sources (\textsc{likelihood}; in {\it orange}), as a function of counts in the 0.2-12 keV and 0.5-7 keV bands, respectively.
    The solid black line shows our chosen threshold.
    }
    \label{fig:likelihood}
\end{figure}

Second, we selected sources based on their detection likelihood.
The likelihood of a source being a real detection and not a background fluctuation is defined as 
\begin{equation}
    L = -\ln(P),
\end{equation}
where $P$ is the probability for a random Poissonian fluctuation to have caused the same observed counts of a real source in the same observation and source aperture.
In both the CSC and 4XMM catalogs, the detection likelihood is computed using the maximum likelihood estimator described in \cite{watson09}:
\begin{equation}
L = -\ln (1 - P_{\Gamma}(\frac{\nu}{2}, \sum_{i=1}^{n} L_i)),
\end{equation}
where $P_{\Gamma}$ represents the incomplete Gamma function, $n$ is the number of energy bands involved, $\nu = 2 + n$ for point-like sources, and $L_i = C_i / 2$, with $C_i$ as defined by \citet{cash79}.
In the CSC catalog, this is captured by the \textsc{likelihood} parameter at the master level, which aggregates data across all stacked observations and energy bands. In 4XMM, instead, this quantity is called \textsc{EP\_*\_DET\_ML} and is computed for each energy band.  We used band 8 (0.2$-$12 keV), which encompasses the sum of the single-band likelihoods across the PN, MOS1, and MOS2 cameras. 
This choice ensures that the detection likelihood metric used is consistent and comparable between the CSC and 4XMM catalogs, facilitating a unified analysis of X-ray source detections across both datasets.
For both the CSC and 4XMM, we selected sources with a conservative combined detection threshold of $L>10$, roughly corresponding to $\sigma \gtrsim 4$. We show the likelihood as a function of net counts (i.e., background subtracted) for all the point-like sources inside the S82-XL area in Figure \ref{fig:likelihood}. This is a lower threshold compared to the \citetalias{lamassa16} S82X catalog ($>$5 and $>$4.5 $\sigma$ for \xmm and \chandra, respectively); however, S82X was focused on targeting bright sources, and several works have shown that with this more permissive threshold the fraction of spurious detections will be limited to $<1$\% (\citealp[e.g.,][]{cappelluti07, civano16, brunner22}).
A summary of detections per energy band is provided in Table \ref{tab:detections}. 

\begin{figure}[!tp]
    \centering \vspace{0.5cm}
    \includegraphics[scale=0.55]{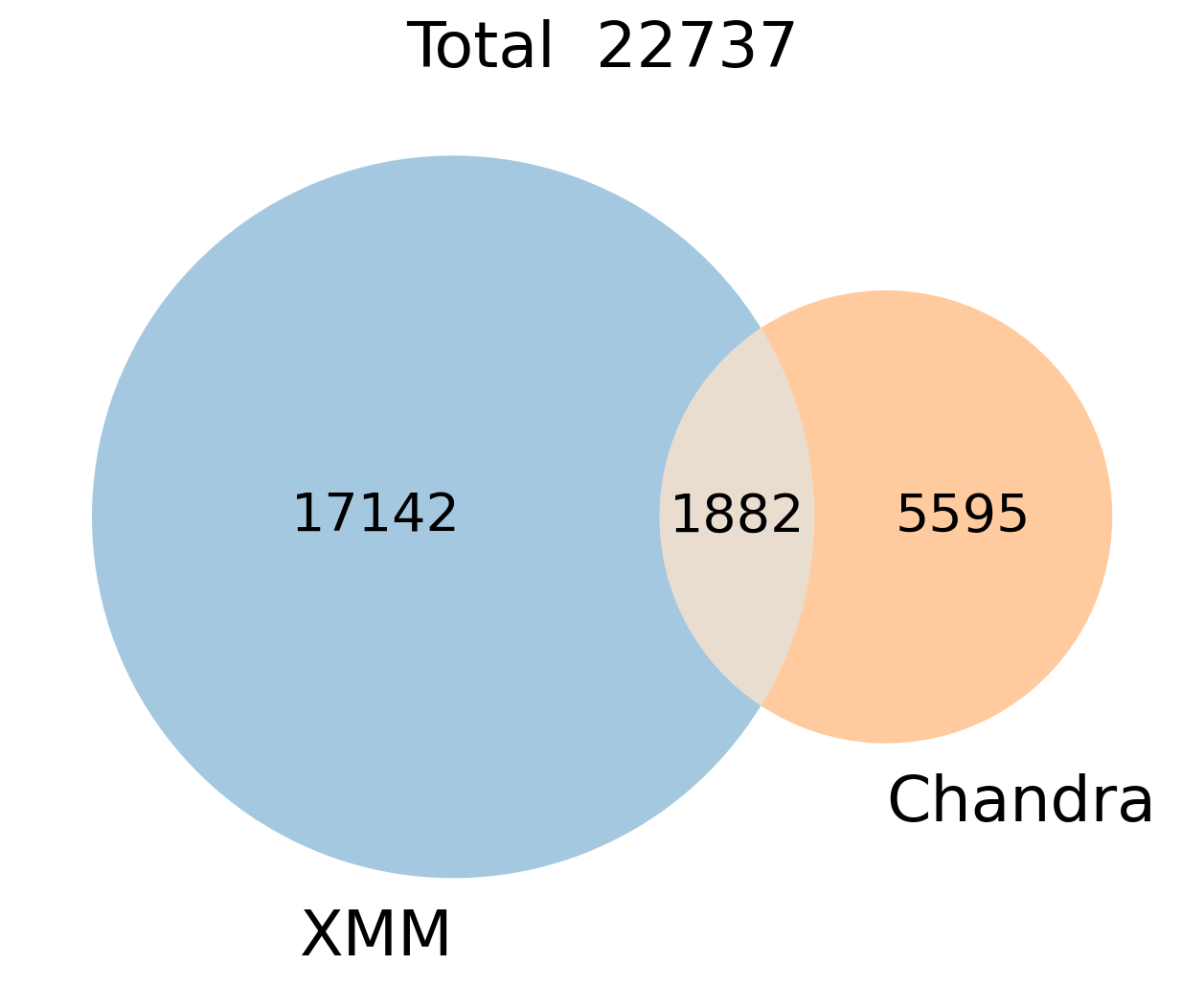}
    \caption{Total number of unique sources detected with a significance of at least $4\sigma$ in the 4XMM (blue) and \chandra CSC (orange) catalogs. Of 22,737 total sources, 1882 are detected by both telescopes.}
    \label{fig:xray-sources_venn}
\end{figure}

\begin{figure}[!tp]
    \includegraphics[scale=0.45]{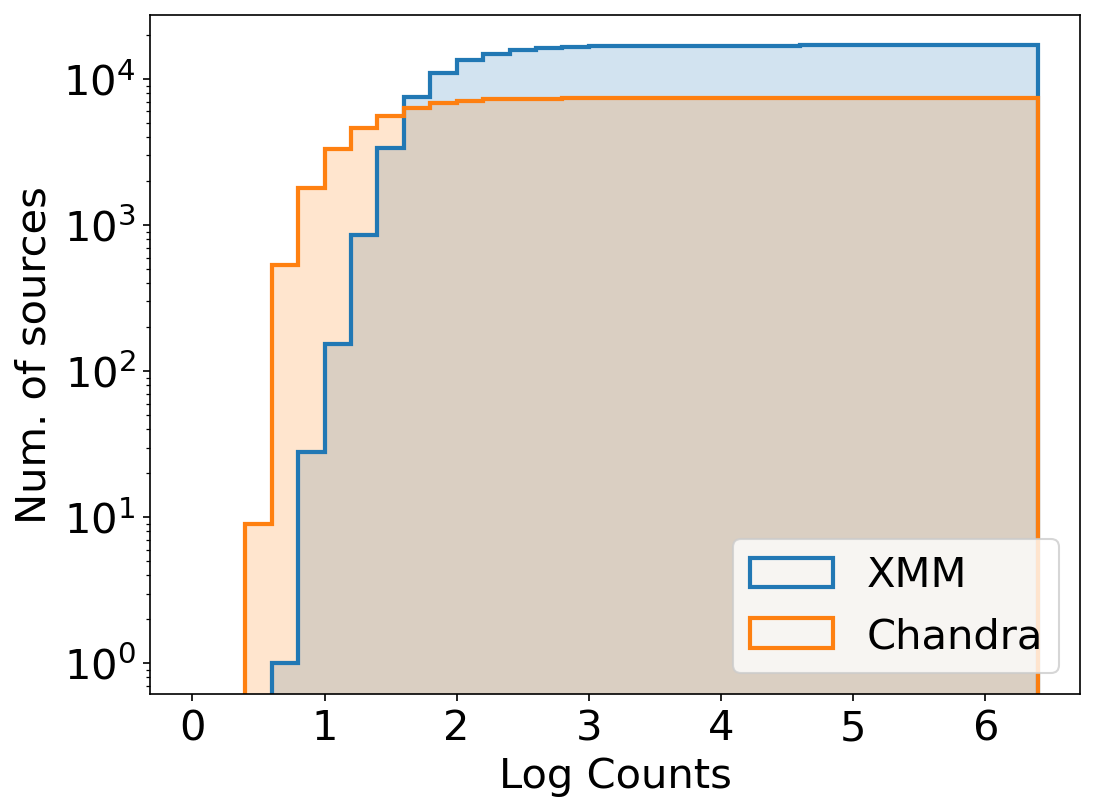}
    \caption{Cumulative number of source counts in S82-XL for \xmm (blue) and \chandra (orange). Here, the number of counts refers to the \xmm (0.2$-$12 keV) and \chandra (0.5$-$7 keV) energy bands reported in the 4XMM and CSC, respectively.}
    \label{fig:num_counts}
\end{figure}

\begin{figure*}[!tp]
    \includegraphics[scale=0.4]{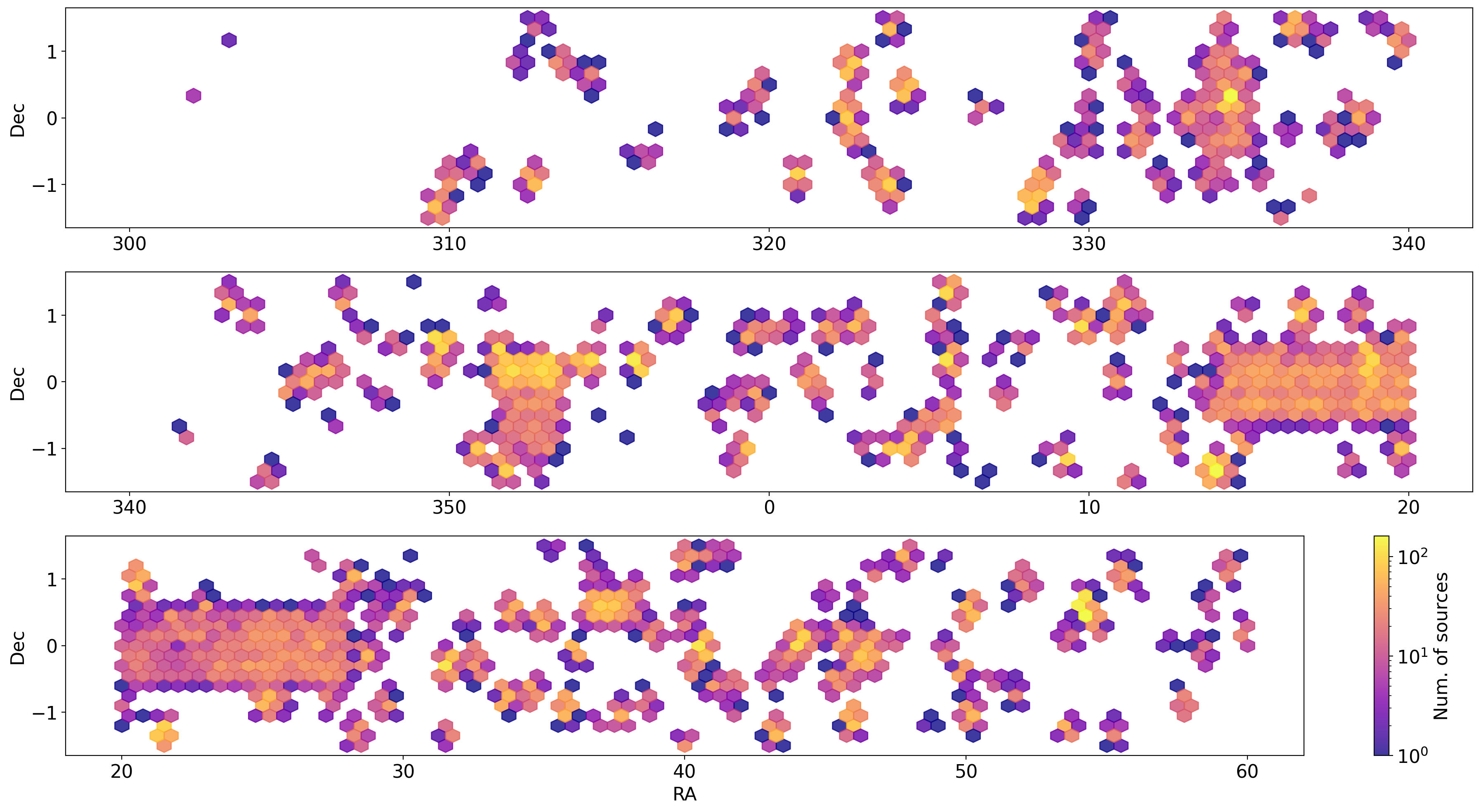}
    \caption{Hexbin density plot showing the number of X-ray sources in the S82-XL region covered by all the available observations (374 by \xmm, and 335 by \chandra). The covered area has been split into three sub-stripes for graphic purposes.}
    \label{fig:sky_map_sources}
\end{figure*}

Some of the \chandra and \xmm observations overlap each other (completely or partially). 
To match \xmm and \chandra sources coming from these overlapping areas, we applied the matching estimator developed by \cite{cappelluti07,cappelluti16}. Where multiple sources are inside the matching radius, chosen to be 15\arcsec\, as in \citetalias{lamassa16},  the best match corresponds to the source that minimizes the following quantity:
\begin{equation}
    R^2 = \left(\frac{\Delta \mathrm{RA}}{\sigma_{\mathrm{RA}}}\right)^2 + \left(\frac{\Delta \mathrm{Dec}}{\sigma_{\mathrm{Dec}}}\right)^2 + \left(\frac{\Delta F}{\sigma_{F}}\right)^2,
\end{equation}
where $\Delta \mathrm{RA}$ and $\Delta \mathrm{Dec}$ are the differences between \chandra and \xmm coordinates, $\Delta F$ is the flux difference in the 0.5$-$10 keV band, and the $\sigma$s represent the largest uncertainties between the two telescopes. 
For \xmm, we preferred the values from the stacked catalog when available, and used the values from the standard catalog otherwise. The two catalogs are cross-matched by the 4XMM team \citep{traulsen20}.
For \chandra, we converted the fluxes from 0.5-7.0 keV to 0.5-10 keV by assuming a power-law with $\Gamma=1.4$ (e.g., \citealp{marchesi16b, nanni20}). This matches the shape of the cosmic X-ray background (\citealp{hickox06}) and, consequently, the average photon index of both obscured and unobscured AGNs.
The resulting X-ray catalog comprises 22,737 X-ray point-like sources, of which 17,142 were detected by \xmm, 5,595 by \chandra, and 1,882 by both telescopes (Figure \ref{fig:xray-sources_venn}). The number of counts and fluxes are shown in Figure \ref{fig:num_counts}, while the source distribution across the S82-XL area is shown in Figure \ref{fig:sky_map_sources}. 

\begin{figure}[!tp]
\vspace{0.2cm}
    \includegraphics[scale=0.44]{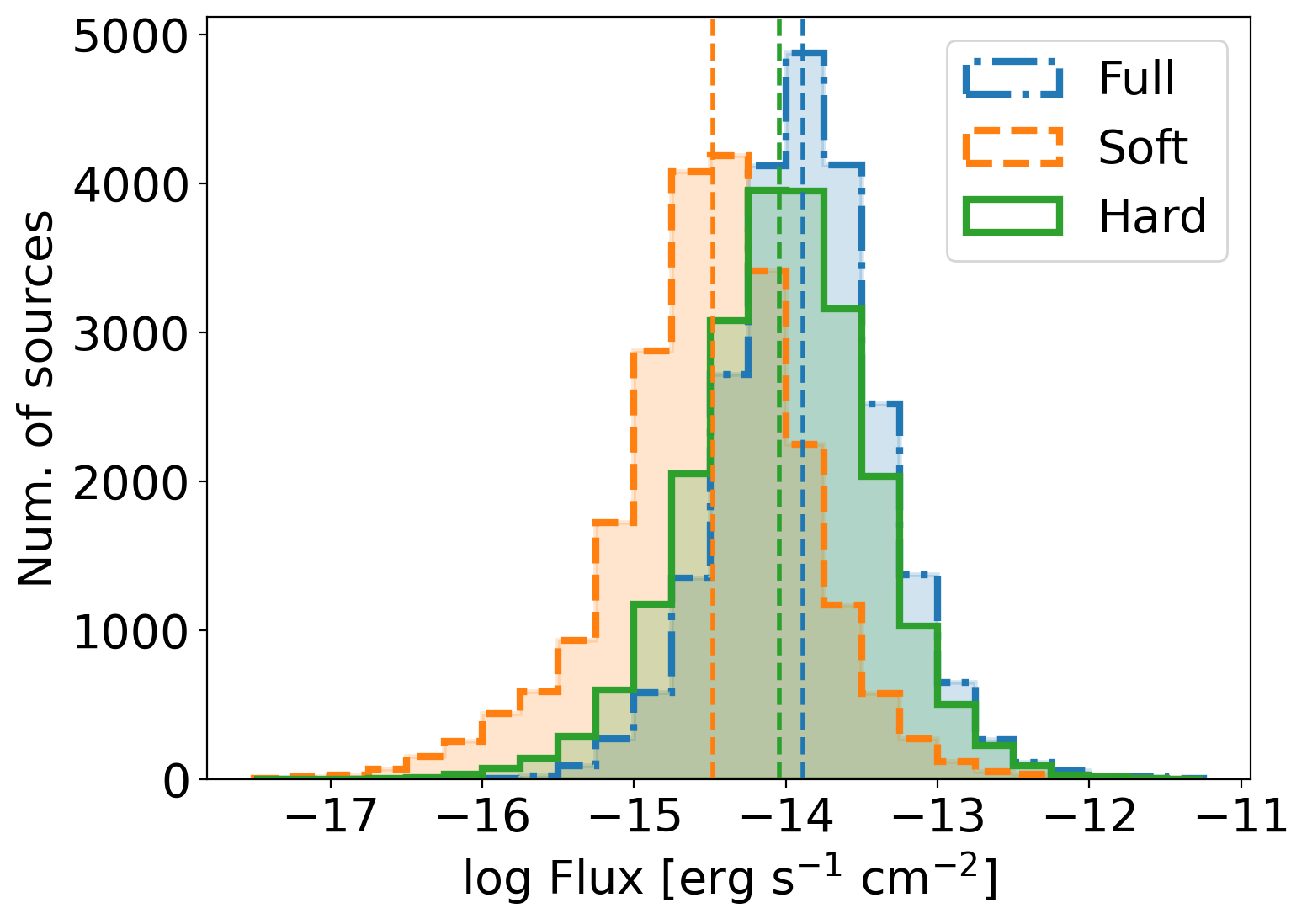}
    \caption{Distributions of fluxes in the full (0.5$-$10 keV; blue, dot-dashed), soft (0.5$-$2 keV; orange, dashed), and hard (2$-$10 keV; green, solid) energy bands for \xmm and \chandra sources. When the source flux was available in more than one catalog, we assumed the following priority order: 4XMMs, 4XMM, and CSC. The dashed lines represent median values.}
    \label{fig:fluxes}
\end{figure}

Figure \ref{fig:fluxes} shows the flux distributions. For \chandra, we used the \textsc{flux\_aper\_avg\_*} columns, which correspond to the aperture-corrected net energy flux inferred from the source region aperture, averaged over all contributing observations, and calculated by counting X-ray events for each science energy band (\citealp{evans10}). For \xmm, we used the \textsc{EP\_*\_Flux} columns, which are the mean band fluxes for the individual cameras obtained by adding the fluxes in each band (\citealp{webb20}).
We standardized the fluxes to the standard full (0.5$-$10 keV), soft (0.5$-$2 keV), and hard (2$-$10 keV) energy bands as follows. For the soft band, we summed together the fluxes in bands 2 and 3 for \xmm and band s and m for \chandra. For the hard band, we summed the fluxes in bands 4 and 5 for \xmm and used band h for \chandra; then, we converted to the desired energy band by assuming a power-law with $\Gamma = 1.4$ and Galactic absorption of $\sim4\times10^{20}$ cm$^{-2}$ (median value in the S82-XL area; \citealp{kalberla05}). 
The same approach was followed for the full band, where we rescaled the \xmm's band 8, and \chandra's band b and w. The derived median fluxes are $1.3_{-0.9}^{+2.7}\times 10^{-14}$, $3.3_{-2.5}^{+8.3}\times 10^{-15}$, $9.1_{-7.1}^{+23.8}\times 10^{-15}$ erg s$^{-1}$ cm$^{-2}$ for the full, soft, and hard bands, respectively, where the uncertainties are the 16th and 84th percentiles of the distributions.

\hspace{-1cm}
\begin{table}[!t]
\hspace{-1.6cm}
\centering
\begin{minipage}[b]{0.35\linewidth}
\centering
\begin{tabular}{ccc}

\hline
\multicolumn{3}{c}{\textbf{Chandra}}    \\
Band   &   keV   & \# Srcs  \\
\hline
u   &  0.2-0.5   & 1088   \\
s   &  0.5-1     & 5159   \\
m   &  1-2       & 6853   \\
h   &  2-7       & 6381   \\
b   &  0.5-7     & 6960   \\
w   &  0.1-10    & 69     \\
\hline
\end{tabular}
\end{minipage}%
\hspace{7mm}
\begin{minipage}[b]{0.35\linewidth}
\centering
\begin{tabular}{ccc}
\hline
\multicolumn{3}{c}{\textbf{XMM-Newton}} \\
Band   &   keV   & \# Srcs  \\
\hline
1   &  0.2-0.5    & 14460 (8783)   \\
2   &  0.5-1      & 14940 (9040)   \\
3   &  1-2        & 15158 (9108)   \\
4   &  2-4.5      & 14787 (8988)   \\
5   &  4.5-12     & 13643 (8564)   \\
8   &  0.2-12     & 15431 (9204)   \\
\hline
\end{tabular}
\end{minipage}
\caption{Number of S82-XL sources per energy band selected from the CSC (\textit{Chandra}, left) and 4XMM (\textit{XMM-Newton}, right). Numbers within parenthesis are sources detected in the stacked 4XMMs catalog.}
\label{tab:detections}
\end{table}

\subsection{Comparison with Stripe 82X DR3 catalog}
We cross-matched the S82-XL with the \citetalias{lamassa16} Stripe 82X Data Release 3 (S82X DR3; \citealp{lamassa24}) master catalog. In principle, CSC and 4XMM should contain all the observations used by \citetalias{lamassa16}, and therefore, also the same number of sources. However, we found 348 sources in S82X that are missing in S82-XL. 148 of them are located in high-background regions of XMM AO13 and AO10 data (see \citealp{lamassa13b} and \citetalias{lamassa16} for a description of these proprietary observations) that were discarded when building the 4XMM catalog but included in \citetalias{lamassa16} by doing a careful source detection. Because of that, we decided to include these sources in S82-XL.
The discrepancy of the remaining 200 ($\sim$3\%) sources can be attributed to the following reasons. Among these, 83 are classified as extended in CSC and/or 4XMM, and 21 have a likelihood close to but below the chosen threshold. These differences likely come from the diverse methodologies employed in source detection across catalogs and from the impact of incorporating new, overlapping observational data, where depth and source variability could also influence the outcome. The absence of the rest of the sources is believed to be due to similar reasons, such as depth, variability, and different data selection strategies and analysis in the source detection process. Furthermore, it is important to note that \citetalias{lamassa16} relied on CSC v1.1, known for its higher rate of false detections compared to the more reliable CSC v2.1\footnote{https://cxc.cfa.harvard.edu/csc/about.html.}. This earlier version was prone to mistaking pile-up artifacts and anomalies near CCD gaps as genuine sources, leading to potential inaccuracies. 
Because of these reasons, we did not include these sources in our S82-XL catalog.

S82-XL includes a variety of newer Chandra and XMM-Newton observations in addition to the original S82X program. These include follow-up observations of interesting sources detected in other bands, investigations of specific source populations, and other targeted studies at low and high fluxes, with exposure times ranging from a few to $\sim$ 100 ks for both \xmm and \chandra\footnote{For additional details on the observations, we refer to the 4XMM and CSC source catalogs.}. Despite this, the overall computed $\log N$-$\log S$, as shown in Section \ref{sec:catalog}, agrees well with other studies, suggesting that any potential biases introduced by these targeted campaigns are mitigated by the diverse range of observations included in the survey.

\section{Sensitivities and number of counts}\label{sec:catalog}
To compute the sensitivities of S82-XL we start from the data products available for the CSC and 4XMM source catalogs.
For \chandra, sensitivity maps are already available for all the observations in the s, m, h, and b bands. These maps were computed by using a detection likelihood classification as ``True" and ``Marginal", corresponding to a false source rate of 0.1 and 1 false sources per observation, respectively. We used the maps with classification as ``Marginal" since it roughly corresponds to the same detection likelihood of 10 we used to select sources\footnote{See details on \href{https://cxc.cfa.harvard.edu/csc/memos/files/Nowak\_csc2\_thresholds.pdf}{https://cxc.cfa.harvard.edu/csc/memos/files/\\Nowak\_csc2\_thresholds.pdf}}.
We used the \textsc{reproject\_image\_grid} CIAO tool to project the sensitivity maps in the common reference system centered at the center of the Stripe 82-XL region (RA=0.0 and Dec=0.0 degrees). We used a grouping factor of 8 to match the binning of the sensitivity maps. Then, we created a mosaic of sensitivity maps using the \textsc{dmimgfilt} CIAO tool with the prescription that when two or more maps contribute to the same image pixel, the map with the lowest pixel values, i.e., with the highest sensitivity, is used. The sensitivity maps are in units of photons s$^{-1}$ cm$^{-2}$, i.e., the photon flux that corresponds to the minimum Poisson background fluctuation in the source aperture that would exceed the source detection likelihood threshold. We converted it to energy flux by applying the calibration suggested by the CSC documentation:
\begin{equation}
    \log F = m \log(\mathrm{photon\, flux}) + c ,
\end{equation}
where $F$ is the flux in erg s$^{-1}$ cm$^{-2}$, and $m$ and $c$ are energy-dependent constants available in the CSC webpage\footnote{https://cxc.cfa.harvard.edu/csc/char.html}. 
Since we are interested in the standard 0.5$-$2, 2$-$10, and 0.5$-$10 keV energy bands, we combined bands s and m, following the procedure described in \cite{lamassa13a} to obtain a sensitivity map in the 0.5$-$2 keV energy range. In short, we created new 0.5$-$2 keV sensitivity maps by adding together background images in band s and m to produce a 0.5$-$2 keV background image. This was then used in the production of a new soft band sensitivity map with the CIAO tool \textsc{lim\_sens}.
The cumulative sky coverage (area vs. sensitivity curves) was finally computed on the resulting image by making a cumulative histogram of all the pixel values. For the 2$-$10 and 0.5$-$10 keV bands we used bands h and b, respectively, and extrapolated assuming a power-law with $\Gamma=1.4$.

For \xmm, instead, the sensitivity maps are available only in the 0.2$-$12 keV band. Therefore, we created our own sensitivity maps as follows. First, we used the SAS tool \textsc{emask} to create detection masks for each energy band, where we used the exposure maps provided by 4XMM. Second, we used the SAS tool \textsc{esensmap} to create sensitivity maps for each 4XMM energy band. We set the likelihood value to 10 to match the minimum detection likelihood used in this work.
In addition to band 8, we combined together background maps for energy bands 2 and 3, and 4 and 5, to get sensitivity maps in 0.5$-$2 keV and 2$-$12 keV energy ranges, respectively, as we did for \chandra. Again, for the 2$-$10 and 0.5$-$10 keV bands we extrapolated assuming a power-law with $\Gamma=1.4$.
Once we obtained the sensitivity maps, we proceeded as we did for \chandra.
We acknowledge that our methodology does not incorporate simulations to evaluate the potential inclusion of spurious sources or to assess the accuracy of our flux measurements, as commonly done in similar studies (\citealp[e.g.,][]{civano16,masini20,brunner22}). This decision was informed by the comprehensive validation already conducted in the creation of the 4XMM and CSC source catalogs. Consequently, we remain confident in the robustness of our approach, deeming it sufficiently reliable without the need for further simulation-based adjustments.

\begin{figure*}[!tp]
    \includegraphics[scale=0.47]{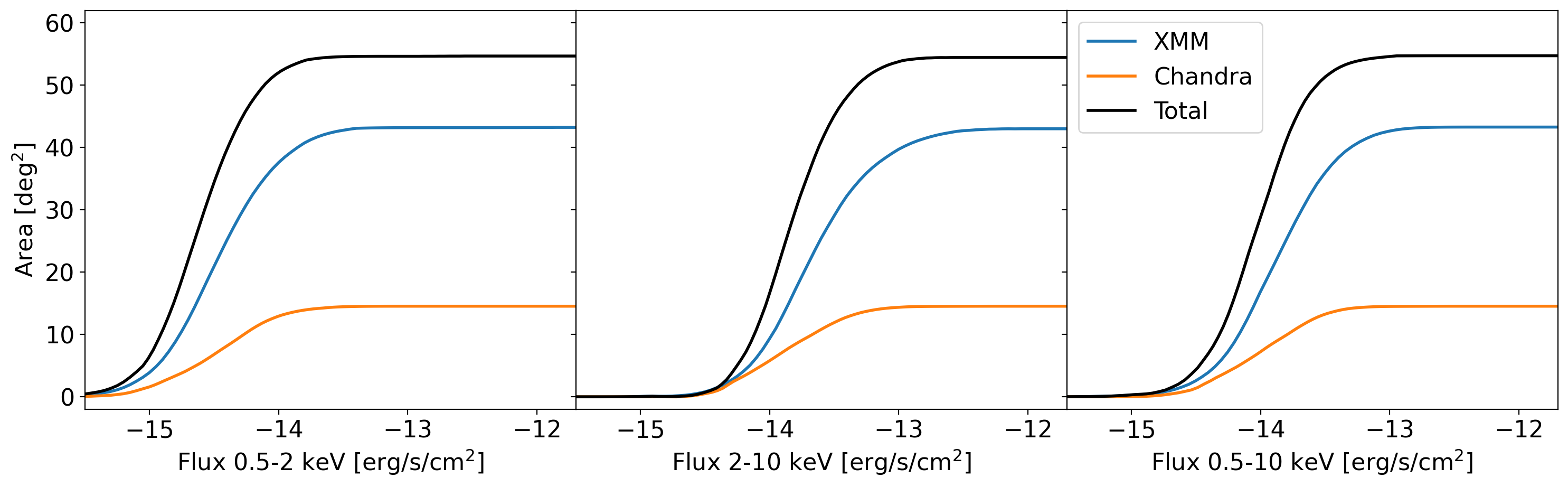}
    \caption{Sky coverage of S82-XL in the soft ({\it left}), hard ({\it middle}), and full ({\it right}) energy bands. The total ({\it black}) curve is the sum of the \xmm ({\it blue}) and non-overlapping \chandra observations. The total \chandra data are also shown ({\it orange}). The S82-XL reaches $\sim54.8$ deg$^2$ of non-overlapping area and limiting fluxes of 3.4$\times 10^{-16}$, 2.9$\times 10^{-15}$, and 1.4$\times 10^{-15}$ erg s$^{-1}$ cm$^{-2}$ in the soft (0.5-2 keV), hard (2-10 keV) and full (0.5-10 keV) bands, respectively}
    \label{fig:sensitivities}
\end{figure*}

\begin{figure*}[!tp]
    \includegraphics[scale=0.43]{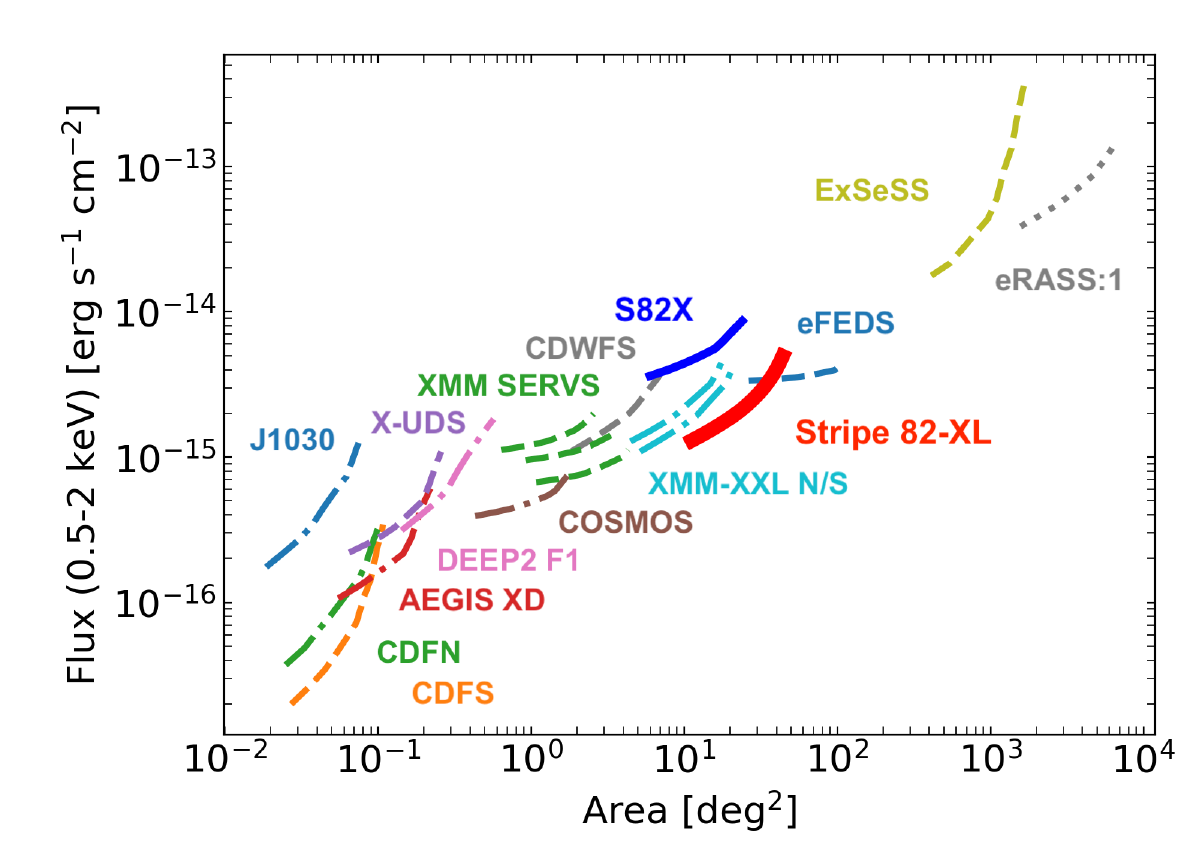}
    \hspace{3mm}
    \includegraphics[scale=0.43]{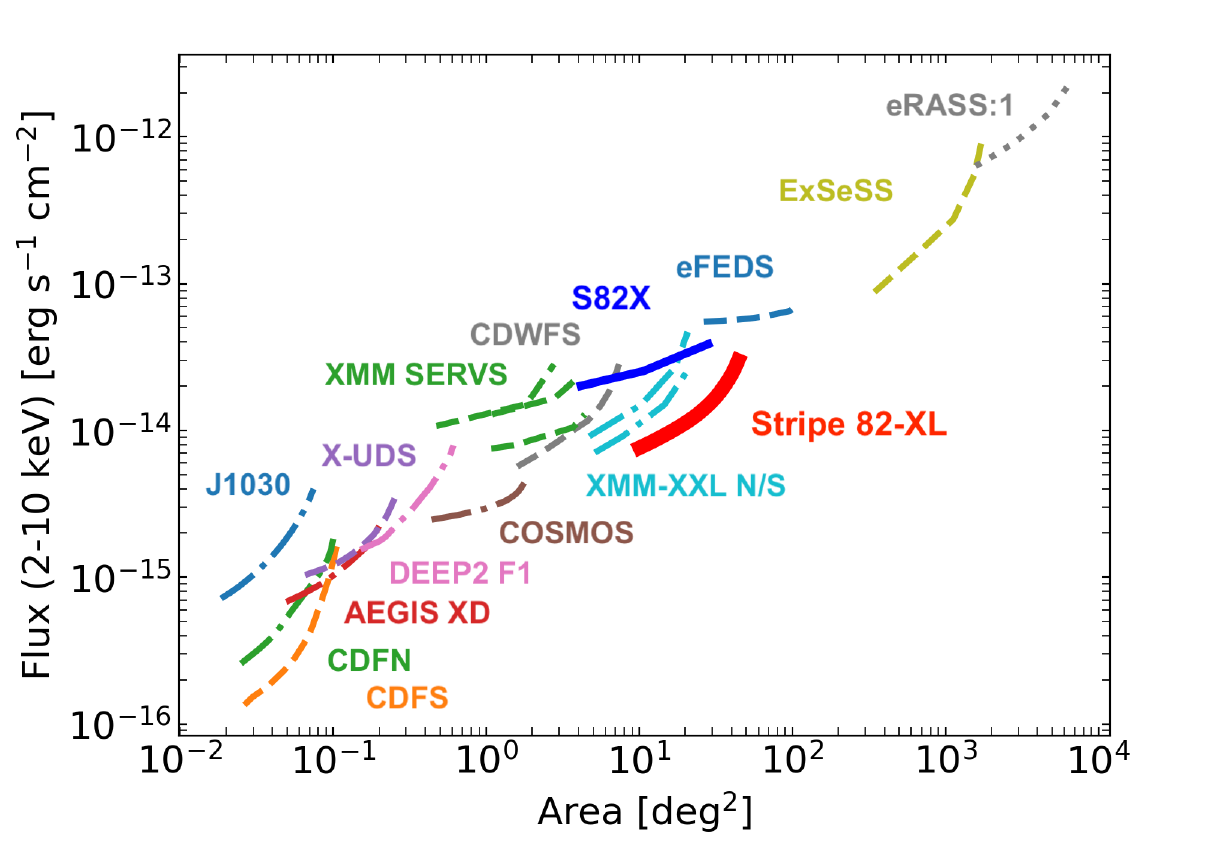}
    \caption{Flux-area sensitivity curves of some of the most relevant surveys in soft (0.5-2 keV, left) and hard (2-10 keV, right) energy bands. All the curves are cut at 20\% and 80\% of the total covered area. From the bottom left: CDFS (orange, \citealp{xue11}), CDFN (green, \citealp{xue16}), AEGIS-XD (red, \citealp{nandra15}), J1030 (blue, \citealp{nanni20}), X-UDS (purple, \citealp{kocevski18}), DEEP2 F1 (pink, \citealp{goulding12}), COSMOS-Legacy (brown, \citealp{civano16}, XMM-SERVS (green, \citealp{chen18,ni21}), CDWFS (grey, \citealp{masini20}), XMM-XXL North and South (cyan, \citealp{chiappetti18}), Stripe 82X (dark blue, \citealp{lamassa16}) and XL (red, this work), eFEDS (blue, \citealp{brunner22}), ExSeSS (yellow, \citealp{delaney23}), and eRASS:1 (grey, \citealp{merloni24}). eROSITA eFEDS and eRASS:1 are estimated from the soft coverage by assuming the ratio between soft and hard sensitivity as shown in \citealp{merloni13}.}
    \label{fig:sensitivities_allsurveys}
\end{figure*}

The resulting area versus sensitivity curves are shown in Figure \ref{fig:sensitivities}. Here, for the total sensitivity, we prioritized \xmm observations due to the larger coverage and deeper exposures, and excluded \chandra observations where there is overlap.
The S82-XL covers $\sim54.8$ deg$^2$ of non-overlapping area, to limiting fluxes\footnote{We refer here to fluxes corresponding to 1\% of the covered area.} of 3.4$\times 10^{-16}$, 2.9$\times 10^{-15}$, and 1.4$\times 10^{-15}$ erg s$^{-1}$ cm$^{-2}$ in the soft (0.5$-$2 keV), hard (2$-$10 keV) and full (0.5$-$10 keV) bands, respectively; the limiting fluxes at half-area survey are 2.4$\times 10^{-15}$, 1.5$\times 10^{-14}$, and 9.5$\times 10^{-15}$ erg s$^{-1}$ cm$^{-2}$ in the soft, hard and full bands, respectively.
Figure \ref{fig:sensitivities_allsurveys} shows the S82-XL contribution to the field of X-rays surveys. By almost doubling the S82X area and improving its limiting fluxes by a factor of $\sim$4 (and half-area sensitivity by a factor of $\sim2$) in all the bands of interest, we have significantly improved the flux-area parameter space covered by X-ray surveys, moving towards lower fluxes and larger areas. Additionally, S82-XL benefits from less stringent detection thresholds, allowing us to fully exploit the X-ray coverage in the region. As a result, S82-XL stands out as one of the most competitive surveys in terms of area and depth.

\begin{figure*}[!tp]
    \centering
    \includegraphics[scale=0.38]{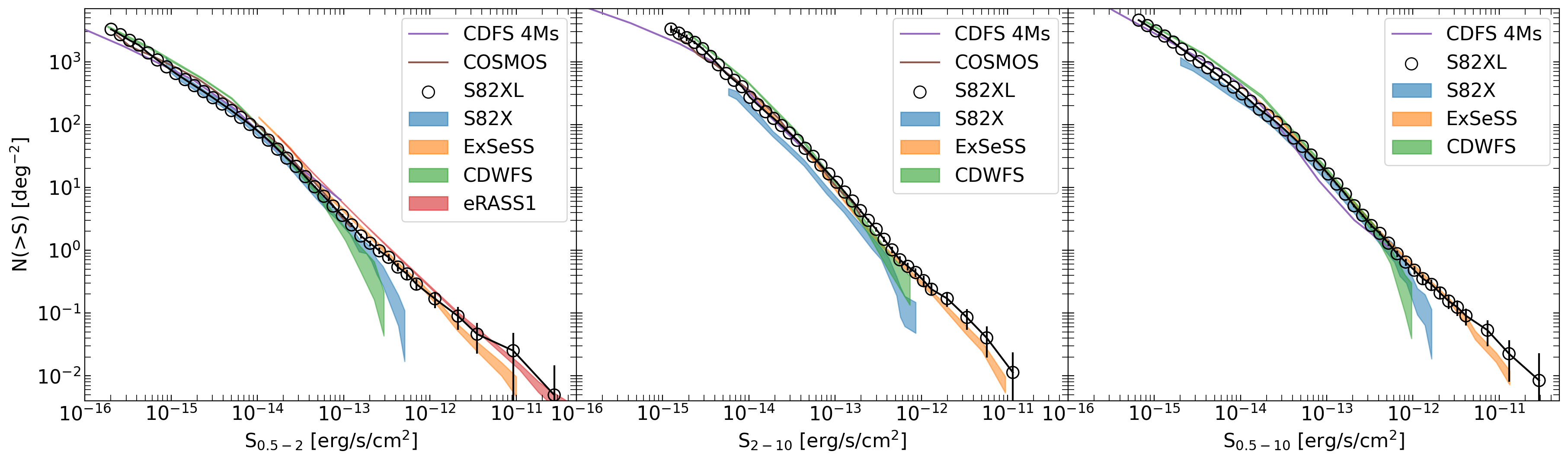}
    \caption{Cumulative number counts for S82-XL ({\it black points}), in the soft ({\it left}), hard ({\it middle}), full ({\it right}) energy bands. Bins are chosen to have at least 8 sources per bin.
    We compare our results to deep, small area surveys CDFS ({\it purple lines}), COSMOS ({\it brown lines}); to similar size surveys S82X ({\it shaded blue}), CDWFS ({\it shaded green}); and to very large area surveys ExSeSS ({\it shaded orange}) and eRASS:1 ({\it shaded red}, soft only).}
    \label{fig:logn_logs}
\end{figure*}

Utilizing the derived sensitivity curves, we calculated the cumulative number counts distribution (i.e., the integral number of sources per unit of flux and area), also known as the $\log N$-$\log S$:
\begin{equation}
    N(>S) = \sum_{i=1}^{N_s} \frac{1}{\Omega_i},
\end{equation}
where $N(>S)$ is the number of sources with flux greater than $S$, $\Omega_i$ is the limiting sky coverage corresponding to the $i$th source, and $N_s$ is the number of sources in each flux bin. The associated uncertainty is the variance,
\begin{equation}
    \sigma^2 = \sum_{i=1}^{N_s} \left( \frac{1}{\Omega_i} \right)^2.
\end{equation}
Figure \ref{fig:logn_logs} shows the $\log N$-$\log S$ across the soft, hard, and full energy bands. 
These derived curves align well with those from other X-ray surveys, maintaining consistency across both the lower and upper flux boundaries. At high fluxes, it is evident how very large-scale ($>50$ deg$^2$) surveys are needed to capture the high end of the source distribution adequately. In contrast, the number densities observed in narrower surveys like CDWFS and S82X tend to diminish significantly at their high flux limits, likely due to a scarcity of sources in these regimes. 
Notably, our data exhibits a marginal increase in number density at high fluxes compared to larger area surveys like ExSeSS and eRASS1. This difference, while consistent within the uncertainties, is likely due to the heterogeneous nature of the survey, which includes targeted observations of bright objects.
At low fluxes, instead, we expect Eddington bias (\citealp{eddington1913})---i.e., the statistical tendency for background fluctuation to scatter more faint sources into a brighter flux bin than vice versa, leading to an overestimation of faint object counts. To mitigate it, we cut our flux distributions where the area curve in each energy band goes below 0.1\% of its maximum value (\citealp{delaney23}). This strategy helps us avoid the less reliable, fainter fluxes at the end of the distributions, thereby indirectly addressing potential biases without direct simulation adjustments.

\section{Observed properties}\label{sec:properties}
\subsection{Redshift luminosity space}
In anticipation of the forthcoming detailed analysis of multiband counterparts and photometric redshifts by Peca et al. (in prep.), we undertook some preliminary work on the S82-XL catalog. Spectroscopic redshifts from the Sloan Digital Sky Survey Data Release 17 (SDSS DR17, \citealp{sdss_dr17}) are already available for \chandra sources in the CSC. For 4XMM we cross-matched with the optical positions of the SDSS spectroscopic catalog, using a matching radius of 7\arcsec\, (as in \citetalias{lamassa16}) and selecting the nearest match as the best candidate. This corresponds to a false association rate of $\sim$4\%\footnote{The fraction of a chance coincidence is estimated as $\rho N_X (\pi r^2) / N_M$, where $\rho$ is the sky density of SDSS sources, $N_X$ the number of X-ray sources, $r$ the matching radius, and $N_M$ the number of matches.}. In line with the procedure used in the CSC, we included only those SDSS spectra flagged with \textsc{ZWARNING=0}, meaning that the spectrum is of sufficient quality to yield a reliable redshift estimation.
However, \citealp{lamassa19_2} showed the limitations of relying solely on the SDSS flag \textsc{ZWARNING=0} as a marker of redshift reliability. Therefore, we conducted a visual inspection of all sources at $z>3$. We focused on this redshift regime since, given the substantial size of the catalog, a visual verification of each source was not feasible. Additionally, possibly wrong estimates at lower redshifts would less significantly affect the catalog due to the large number of sources in these ranges. This effort led to the exclusion of seven redshifts due to corruption or excessive noise in the SDSS spectrum. 

The spectroscopic redshifts from S82X DR3 were also included. For sources with conflicting redshifts, we prioritized the S82X redshifts, for which a more accurate association technique was used. In particular, 187 redshifts derived from our cross-match with SDSS were excluded since the corresponding X-ray sources in S82X lack assigned redshifts. This is likely due to the more accurate maximum likelihood matching technique from \citetalias{lamassa16} and \citealp{ananna17}, which rejected these SDSS sources as X-ray counterparts.  
We obtained final spectroscopic redshift estimates for 6,695 objects, achieving a spectroscopic completeness of $\sim30$\%.

\begin{figure*}[!tp]
    \centering
    \includegraphics[scale=0.47]{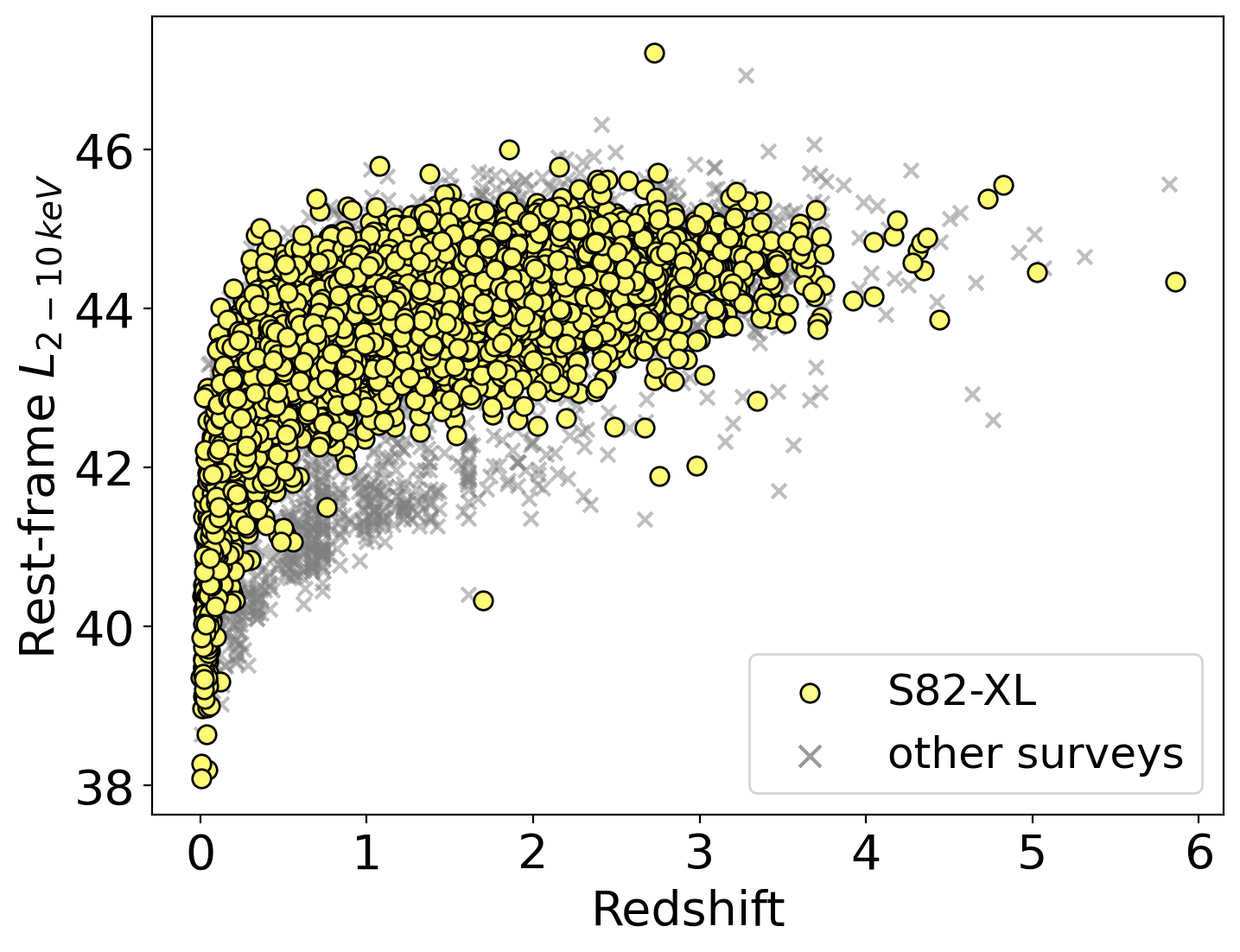}
    \includegraphics[scale=0.47]{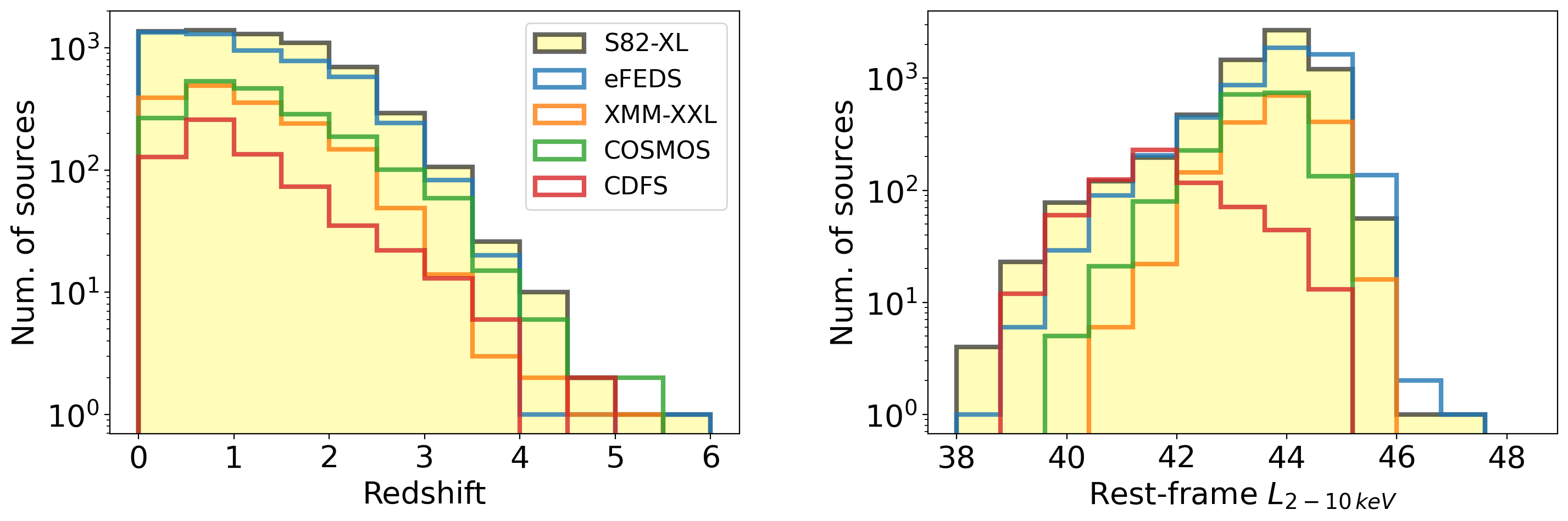}
    \caption{\textit{Top:} Rest-frame, 2-10 keV X-ray luminosity as a function of redshift for S82-XL (yellow points) and other X-ray surveys (grey crosses; eFEDS, XMM-XXL, COSMOS, and CDFS). 
    \textit{Bottom:} Histograms of the redshifts (\textit{left}) and luminosities (\textit{right}) used in the upper plot. Here, we show S82-XL (\textit{yelow}) and other X-ray surveys in color code: eFEDS (\textit{blue}), XMM-XXL (\textit{orange}), COSMOS (\textit{green}), and CDFS (\textit{red}). 
    Note that for eFEDS the hard band luminosity is extrapolated from the 0.2-2.3 keV flux by assuming a power-law with $\Gamma=1.4$. Only sources with spectroscopic redshifts are considered, and luminosities are estimated from observed fluxes (i.e., uncorrected for absorption) in the rest-frame 2-10 keV band.
    S82-XL covers much of the redshift-luminosity plane, more than other X-ray surveys, especially at high redshift and high luminosity.
    }
    \label{fig:zlum_plane}
\end{figure*}

These redshifts allowed us to calculate the rest-frame X-ray luminosities for S82-XL objects. For each band, luminosities were derived with $L_X = F_X \times 4 \pi d_{L}^2 \times k$, where $F_X$ is the flux in the given band, $d_{L}$ is the luminosity distance, and $k=(1+z)^{\Gamma-2}$ is the k-correction to move from observed to rest-frame values, where $\Gamma=1.4$ as assumed throughout this work.
We obtain median values of $1.2_{-1.0}^{+3.4}\times 10^{44}$, $3.7_{-3.4}^{+11.2}\times 10^{43}$, and $7.2_{-6.4}^{+2.6}\times 10^{43}$ erg s$^{-1}$ in the full, soft, and hard bands, respectively, where the uncertainties are computed as the 16th and 84th percentiles of the corresponding luminosity distributions.
It is important to note that these luminosities have not been corrected for potential absorption, meaning they might be influenced by obscuration. While a detailed spectral analysis to determine obscuration would be valuable, it falls outside the scope of this paper and presents an opportunity for future research. Figure \ref{fig:zlum_plane} shows the redshift-luminosity distribution of S82-XL in comparison with other X-ray surveys.
Specifically in the hard band (2-10 keV), the S82-XL catalog presents a wide luminosity range that spans from $1.2\times 10^{38}$ to $1.6\times 10^{47}$ erg s$^{-1}$, reflecting the catalog's capacity to include an extensive array of different X-ray sources. Moreover, the inclusion of spectroscopic redshifts up to $z\sim6$ offers a view of the Universe across early epochs, enhancing the catalog utility. 
Remarkably, S82-XL alone spans the luminosity-redshift space encompassed by many X-ray surveys combined, underlining its unique contribution to the field. 
Large and deep surveys have historically enabled the study of AGN evolution across the parameters of luminosity, absorption, and redshift (\citealp[e.g.,][]{ueda14,buchner14, georgakakis17, ananna19, peca23}). S82-XL extends this capability, allowing for the exploration of these parameters for a broader and more diverse set of AGNs together, facilitating a deeper understanding of AGN evolution and characterization.
We also note that this redshift space will be further covered when photometric redshifts are computed (Peca et al. in prep.), and the spectroscopic redshift campaign started with S82X (\citealp{lamassa16,lamassa19, lamassa24}) will be further enhanced.


\subsection{Hardness ratios and obscuration}\label{subsec:hr}
\begin{figure}[!tp]
    \centering\hspace{-3mm}
    \includegraphics[scale=0.46]{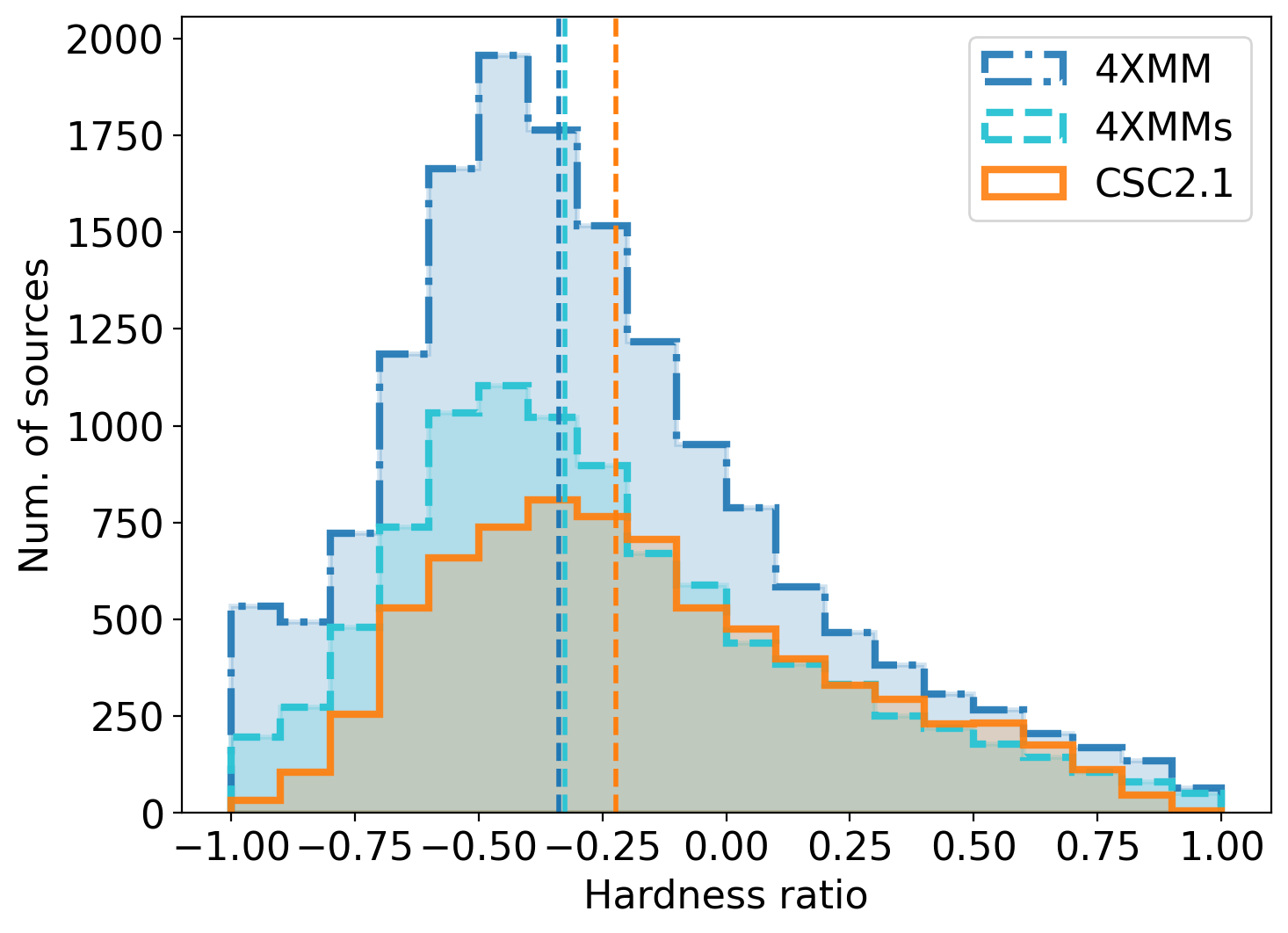}
    \caption{Hardness ratio distributions for all the 4XMM (blue, dot-dashed), 4XMMs (cyan, dashed), and CSC (orange, solid) S82-XL sources. Median values are shown with dashed lines.}
    \label{fig:HR1}
\end{figure}

Hardness ratios (HRs) offer insights into AGN properties, especially regarding obscuration and variability (\citealp[e.g.,][]{marchesi16b, silver23, cox23}). Despite their utility, HR analysis may be biased, influenced by instrumental responses and the spectral shape of the models that needs to be assumed (\citealp[e.g.,][]{masini20,peca21}). However, they offer a quick way to have a first look at the sources and find possible interesting candidates even in the case of low-photon statistics, where a spectral analysis might not be reliable (\citealp[e.g.][]{peca21, peca23}). 
HRs were computed for 4XMM, 4XMMs, and CSC sources, using
\begin{equation}
    HR = \frac{H-S}{H+S} \,,
\end{equation}
where $H$ and $S$ are the hard and soft band count rates, respectively. We derived the count rates using the available bands from 4XMM and CSC source catalogs. For the soft band, we summed the count rates in bands 2 and 3 for \xmm, and \chandra bands s and m to get the count rates in the standard 0.5$-$2 keV band; while for the hard band, we summed values in band 4 and 5 to get the 2$-$12 keV hard band for \xmm and used the h band (2$-$7 keV) for \chandra. Uncertainties are computed using the numerical error-propagation technique outlined in Section 1.7.3 of \cite{lyons1991}. This approach bypasses the limitations of the conventional approximate variance formula (\citealp[e.g.,][]{vignali03,xue11}), particularly when the count numbers are low and the error distribution deviates from Gaussian normality (e.g., Section 2.4.5 of \citealp{eadie1971}).
Band count rates upper limits are used when the nominal values are not available.
The HR distributions for all S82-XL sources are shown in Figure \ref{fig:HR1}. We obtained a median HR of -0.22 for CSC, and -0.34 and -0.33 for 4XMM and 4XMMs sources.
In general, these negative HRs, especially at low redshift since the majority of our sources are within $z\sim 2$, indicate that most of the S82-XL sources are not, or slightly to moderately obscured (\citealp[e.g.,][]{tozzi01, peca21} and references therein). 

\begin{figure}[!tp]
    \centering\hspace{-5mm}
    \includegraphics[scale=0.57]{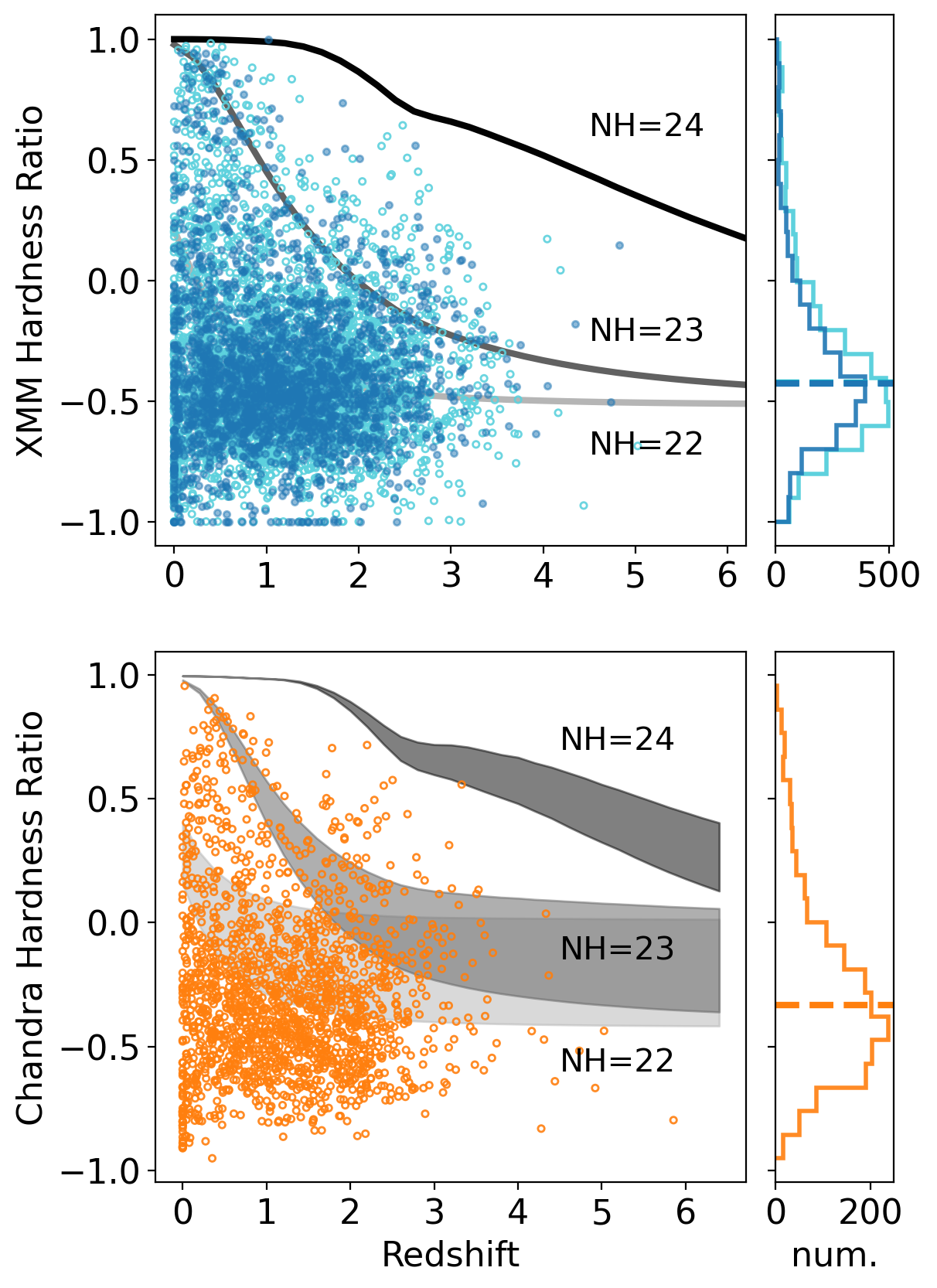}
    \caption{Hardness ratios for 4XMM (top, blue), 4XMMs (top, cyan), and CSC (bottom, orange) as a function of redshift for S82-XL sources. Hardness ratio distributions are also plotted on the right of each panel, as well as median values (dashed lines). If the HR is available for both 4XMM and 4XMMs, we show the 4XMMs value. We also show the expected curves when assuming a single, absorbed power-law with $\Gamma=1.8$, and absorption $\log N_H/\mathrm{cm}^{-2}=$ 22, 23, and 24 (color-coded in greyscale from lighter to darker, respectively). While the \xmm curves are computed by considering the average response matrices of the selected observations, the width of \chandra curves is computed by considering the earliest and latest response matrices for S82-XL sources (see text for details).}
    \label{fig:HR2}
\end{figure}

We used the \xmm and \chandra response matrices to predict expected HR trends as a function of $N_H$ and redshift, assuming a simple power-law model with a standard $\Gamma=1.8$ (\citealp[e.g.,]{nandra94,piconcelli05}), modified by Galactic (\citealp{kalberla05}) and rest-frame absorption $N_H$. We used \textsc{PyXSPEC} (\citealp{pyxspec, xspec}) to simulate the predicted values. 
Given that S82-XL incorporates data spanning from the earliest \xmm and \chandra cycles up to 2021, we adapted our analysis to account for variations in instrumental efficiency. 
For \xmm, for which we found that its effective area has remained relatively stable between revolution 27 and 3843 (the oldest and newest observations, respectively), we used average HR trends obtained from all observation cycles. For Chandra, instead, it is not possible to consider unique HR trends due to the known degradation in the soft band (e.g., \citealp{peca21}). As detailed in Appendix \ref{app:chandra_eff}, this degradation leads to fluctuations in the expected HR across different cycles. We decided to create HR trends as a function of different Chandra cycles (specifically, cycles 14, 18, 19, 20, 21, and 22) where the variation in the soft-to-hard effective area ratio exceeds $\sim0.5$ dex. For $\log N_H/\mathrm{cm}^{-2} < 22$, where the variation most significantly affects the hardness of a single absorbed power-law, we achieve an HR precision of $\lesssim 0.15$ between the different cycles. This precision falls within the typical HR uncertainty for our sources. 
Expected HR trends are also created for both \chandra ACIS-S and ACIS-I cameras separately. Using these specified response matrices, we constructed a grid of expected values (e.g., \citealp{masini20}) and derived the $N_H$ for sources with spectroscopic redshift (Figure \ref{fig:HR2}). When HR values are available for both \xmm and \chandra, we used the values coming from the observation with the higher exposure time.
We obtained a median HR of -0.33 and -0.43 for \chandra and \xmm, respectively. 
These sources, for which a spectroscopic redshift is available, are softer than the whole sample (see Figure \ref{fig:HR1}). This is because hard sources are generally more obscured, and therefore fainter, making spectroscopic optical surveys negatively biased against them.
Even if the hard band used for \xmm is more extended, the \chandra HRs are more positive, possibly due to the differences in the effective areas of the two telescopes. 

For the column density, we obtained a median value of $\log N_H/\mathrm{cm^{-2}}=21.6_{-1.6}^{+1.0}$, and an overall, obscured fraction ($\log N_H/\mathrm{cm^{-2}}>22$) of $\sim 36.9\%$. Comparing these results with surveys employing similar methods for computing $N_H$, we find our obscured fraction to be close to that of COSMOS ($\sim 37.1\%)$ and lower than that of the CDFS ($\sim 45.9\%$), possibly due to the deeper fluxes reached by CDFS (see also Figure \ref{fig:zlum_plane}).
Employing a more complex and physically motivated spectral model (e.g., \texttt{MYTorus} \citealp{murphy09}, \texttt{borus02} \citealp{balokovic18}, \texttt{UXCLUMPY} \citealp{buchner19}) could potentially offer a more accurate $N_H$ estimation for these sources. However, such an approach might also complicate the interpretation of our data by causing the predicted HR tracks to overlap each other. This overlap could introduce further ambiguity in assigning a $N_H$ value to a source with a specific HR and redshift (\citealp{peca21}). Therefore, we have chosen not to pursue a more complex modeling in this analysis, keeping in mind that our primary aim is to illustrate the general hardness of the sources in the S82-XL catalog rather than to draw strong conclusions regarding their obscuration. It is also worth mentioning, it should be noted that relying solely on a single power-law model could result in the underestimation of some heavily obscured AGNs, that might exhibit softer HRs because of an extra component in the soft X-rays, typically attributed to the leakage or scattering of the primary power-law component (\citealp{lambrides2020,masini20,peca23, peca23axis}).

\section{Summary and conclusions}\label{sec:summary}
In this paper, we introduced the Stripe 82 XL (S82-XL) catalog, a significant enhancement of the S82X survey. 
The catalog is described in Table \ref{tab:data_columns} in Appendix \ref{app:cat_description}.
Here are the key takeaways from our study:

\begin{itemize}
    \item By leveraging the \xmm and \chandra source catalogs, we compiled a rich dataset of X-ray sources from observations taken up until December 31, 2021. The final catalog features 22,737 unique point-like sources, covering a total, non-overlapping area of $\sim$54.8 deg$^2$.
    \item S82-XL significantly improves the flux-area parameter space covered by X-ray surveys (Figure \ref{fig:sensitivities_allsurveys}). This advancement moves towards lower fluxes and larger areas, establishing S82-XL as a pivotal contribution to the field of extragalactic studies.
    \item As shown in Figure \ref{fig:logn_logs}, the derived $\log N$-$\log S$ distributions across the soft, hard, and full energy bands agree with the trends observed in various other X-ray surveys, ranging from deep and narrow to broad and shallow fields. Our analysis emphasizes the importance of large-area surveys in adequately sampling X-ray sources up to high fluxes.
    \item Using redshifts from SDSS ad S82X DR3, we calculated X-ray luminosities, which span nine decades in luminosity and reach redshifts $z\sim6$ (Figure \ref{fig:zlum_plane}). Remarkably, S82-XL covers the luminosity-redshift space to an extent that many other surveys combined do, highlighting its unique contribution.
    \item The hardness ratios (HRs; Figure \ref{fig:HR1} and \ref{fig:HR2}) correspond to relatively low levels of obscuration, as expected in a fairly bright catalog. HRs depend on a combination of source brightness (unobscured AGNs are easier to detect at large distances) and spectral nature of AGNs (whether they are significantly obscured). This underscores the complexity of inferring obscuration levels from HRs alone and highlights the potential for future detailed spectral studies. Despite these caveats, we computed $N_H$ from HRs and obtained an overall obscured fraction of sources of $\sim$36\%.
\end{itemize}

Looking ahead, the S82-XL catalog is set to become a foundational resource for future astronomical research, providing a detailed platform for studying AGN phenomena. The forthcoming multiwavelength counterpart paper (Peca et al. in prep.) will further enhance the S82-XL catalog by extending its coverage from the UV to the radio bands, offering a wealth of multiband information and photometric redshifts. This expansion will enable a more detailed understanding of AGN properties and their environments, facilitating comprehensive studies across the electromagnetic spectrum. 
Additionally, the value of the catalog will be enhanced by the ongoing campaign led by our group (\citealp{lamassa16,lamassa19,lamassa24}), which aims to increase the spectroscopic redshift completeness of the survey. This effort will enrich the S82-XL catalog, allowing for more accurate luminosity calculations and improved constraints on AGN characterization. Together, these enhancements will solidify the role of S82-XL in advancing our understanding of the nature of AGN and their cosmic evolution with unprecedented detail.

\begin{acknowledgements}
We acknowledge the anonymous referee for the valuable comments that improved the quality of the paper.
AP and NC thank the \chandra grant AR2-23010X. AP thanks the CXC and XMM help desks for the help with the CSC and 4XMM source catalogs, in particular R. Marinez-Galarza and E. Traulsen. AP acknowledges D. Costanzo for all the support over the years. 
TTA acknowledges support from ADAP grant number 80NSSC23K0557. 
ET acknowledges support from: ANID through the Millennium Science Initiative Program NCN19\_058, CATA-BASAL ACE210002, and FB210003, and FONDECYT Regular 1190818 and 1200495.
NTA acknowledges support from NASA grants 80NSSC23K0484 and 80NSSC23K1611.
\end{acknowledgements}

\vspace{5mm}
\facilities{\xmm, \chandra}

\software{Astropy v5.0 \citep{astropy:2013,astropy:2018,astropy:2022};
          Pandas v1.4.1 \citep{pandas};
          CIAO v4.15 \citep{ciao_paper},
          SAS v21.0.0 \citep{SAS},
          Matplotlib v3.4.3 \citep{matplotlib},
          TOPCAT v4.94.9 \citep{topcat},
          PyXSPEC v2.1.0 \citep{pyxspec,xspec}
          }

\appendix

\section{Chandra effective area variations in S82-XL}\label{app:chandra_eff}

In this appendix, we briefly discuss the variations in \chandra's effective area over time, to address how these changes impact the reliability of hardness ratio (HR) calculations. This concern arises from the known degradation in \chandra's soft band sensitivity, which can skew HR measurements as the photon collection efficiency decreases more in the soft band than in the hard band (see \citealp{peca21} and references therein for further details\footnote{See also che the Chandra Proposers’ Observatory Guide.}).

\begin{figure}[!bp]
    \centering
    \includegraphics[scale=0.46]{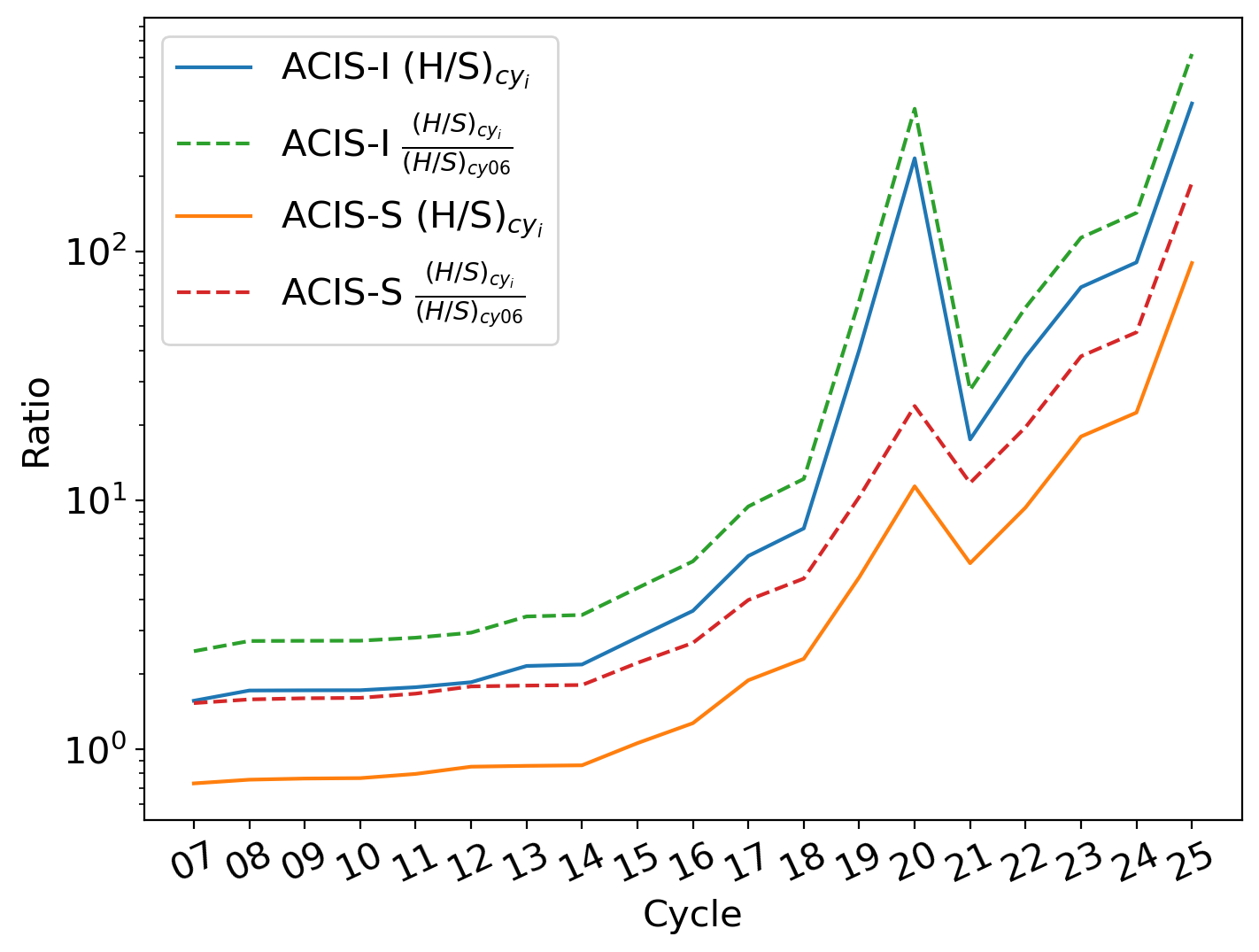}
    \caption{Evolution of the \chandra effective area across different cycles, parametrized with the ratio of hard to soft band effective areas. Each point represents the effective hard-to-soft area ratio of each cycle ($continuous$ lines) and its ratio with the cycle 6 ratio (
    $dotted$ lines), illustrating the gradual degradation in the soft band relative to the hard band. This evolution is shown separately for \chandra ACIS-I ($blue$ and $green$) and ACIS-S ($orange$ and $red$). These variations are important for understanding the impact on hardness ratio calculations.}
    \label{fig:chandra_arf}
\end{figure}

To assess the evolution of the effective area, we examined the ratios between the hard (2-7 keV) and soft (0.5-2 keV) bands for each observational cycle by analyzing the auxiliary response files (ARFs) for both ACIS-I and ACIS-S cameras. In Figure \ref{fig:chandra_arf} we specifically traced the effective hard-to-soft area ratios of each cycle, $(H/S)_{cy_i}$, as well as its evolution over the cycle 6 ratio, $\frac{(H/S)_{cy_i}}{(H/S)_{cy06}}$, which is the oldest cycle for which responses are available. Note that cycle 6 can be considered as a baseline since the effective area prior to this cycle was relatively stable\footnote{CXC private communication.}.

Our analysis revealed variations up to more than two orders of magnitude in the relative difference between the hard and soft bands across different cycles (Figure \ref{fig:chandra_arf}). Based on these findings, we adopted a threshold of $\sim$0.5 dex to define different epochs from which computing HR trends. This led to the selection of response matrices (ARF and RMF) from cycle 14 for all observations prior to this cycle, and subsequently those from cycles 18 through 22, the latter being the most recent in S82-XL. We then used these six effective area epochs to compute expected HR trends as a function of redshift and $N_H$, as detailed in Section \ref{subsec:hr}. Briefly, we assumed a single absorbed power-law model with a standard $\Gamma=1.8$. We opted for this simple model because more complex models tend to introduce degeneracies in the HR maps, complicating the interpretation without significantly improving $N_H$ estimates (e.g., \citealp{masini20,peca21}).
By assuming this model, we observe that the higher HR variations across different \textit{Chandra} cycles are predominantly at lower column densities. Specifically, using the selected responses, we achieve an HR precision of $\lesssim 0.15$ for $\log N_H/\mathrm{cm}^{-2} < 22$, averaging over all redshifts in our sample. This precision is maintained across different cycles and falls within the typical HR uncertainty for our sources. Furthermore, for $\log N_H/\mathrm{cm}^{-2} > 22$, the HR variation between different cycles decreases. Therefore, conducting a more detailed HR analysis with more \chandra cycle responses would not necessarily yield better results, given the high precision relative to the inherent uncertainties in our source measurements. In any case, we stress that while HR analysis provides valuable insights, a detailed spectral analysis remains the preferred and most reliable method for deriving accurate physical AGN properties.

\begin{figure}[!tp]
    \centering
    \includegraphics[scale=0.46]{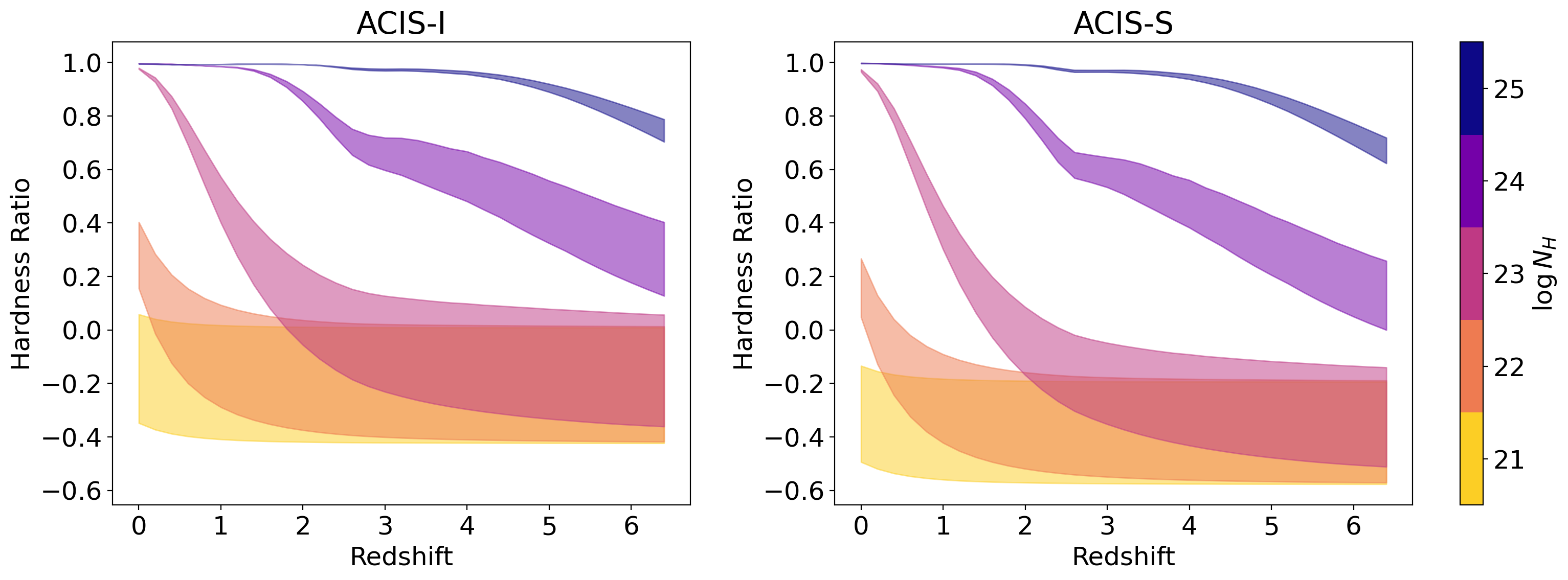}
    \caption{Impact of the effective area changes on the hardness ratio (HR) for \chandra's ACIS-I ($left$) and ACIS-S ($right$) cameras across different cycles in the S82-XL data. The HR values are calculated using a single absorbed power-law model with a photon index $\Gamma=1.8$, Galactic absorption, and varying rest-frame absorption represented in different colors. The width of the curves represents the values obtained using response matrices from the earliest (cycle 6) and more recent (cycle 22) observations in S82-XL, demonstrating how instrumental aging can influence HR measurements and subsequent physical interpretations of AGN properties.}
    \label{fig:acis_hrs}
\end{figure}

For illustrative purposes, Figure \ref{fig:acis_hrs} shows the potential impact on HR calculations across different cycles for ACIS-I and ACIS-S. The responses from cycle 6 and cycle 22 (the oldest and newest observations, respectively) are used to show the varying HRs from observations originating from early cycles, producing lower (i.e., softer) HRs, and more recent cycles, exhibiting higher (i.e., harder) HRs due to the loss of collecting efficiency in the soft band during the years.
These insights highlight the need to adjust analysis techniques to accommodate instrumental variations when dealing with large datasets that span multiple years.

\vspace{0.2cm}
\section{Catalog description}\label{app:cat_description}
The entire S82-XL catalog is hosted on VizieR\footnote{\href{https://vizier.cds.unistra.fr/}{https://vizier.cds.unistra.fr/}} and GitHub\footnote{\href{https://github.com/alessandropeca/S82X}{https://github.com/alessandropeca/S82X}}. The catalog is described in Table \ref{tab:data_columns} below, where errors are reported at the 1$\sigma$ confidence level.

\begin{longtable}{lp{13.6cm}}
\caption{Column description of the S82-XL catalog.\label{tab:data_columns}} \\
\toprule
\textbf{Column Name} & \textbf{Description} \\
\hline
\endfirsthead
\hline
\textbf{Column Name} & \textbf{Description} \\
\hline
\endhead
\hline
\endfoot
\endlastfoot
XID\_S82XL & Unique identifier for sources in the Stripe 82 XL survey. IDs are given by distance from the center of the survey in (RA, Dec) = (0.0, 0.0). \\
RA, DEC, POS\_ERR & Right Ascension and Declination in degrees, representing the celestial coordinates of the sources, and positional error in arcseconds, respectively. When a source is in more than one catalog, we follow the priority: 4XMMs, 4XMM, and CSC. \\
IAUNAME\_4XMM & Source identifier in the 4XMM catalog. \\
OBS\_ID\_4XMM & Observation ID for the XMM observations in 4XMM catalog. \\
RA\_4XMM, DEC\_4XMM & Right Ascension and Declination in the 4XMM catalog. \\
EP\_*\_FLUX\_4XMM & Flux measurements in various energy bands from the 4XMM catalog, with associated errors (EP\_*\_FLUX\_ERR\_4XMM). \\
EP\_8\_CTS\_4XMM & 0.2-12 keV total net counts from the 4XMM catalog, with associated errors (EP\_8\_CTS\_ERR\_4XMM). \\
EP\_*\_RATE\_4XMM & Count rates [cts s$^{-1}$] in various energy bands from the 4XMM catalog, with associated errors (EP\_*\_RATE\_ERR\_4XMM). \\
HR\_4XMM & Hardness ratio calculated using 4XMM data, with associated error (HR\_4XMM\_ERR) computed using standard error propagation. \\
4XMMs & Same columns as [5-32] for the 4XMMs catalog. \\
name\_CSC2p1 & Source identifier in the CSC 2.1 catalog. \\
ra\_CSC2p1, dec\_CSC2p1 & Right Ascension and Declination in the CSC 2.1 catalog. \\
obsid\_CSC2p1 & Observation ID for the Chandra observations in the CSC catalog. \\
flux\_aper\_avg\_*\_CSC2p1 & Flux measurements [erg s$^{-1}$ cm$^{-2}$] in various energy bands from the CSC catalog, with associated lower and upper limits (flux\_aper\_avg\_lolim\_*\_CSC2p1, flux\_aper\_avg\_hilim\_*\_CSC2p1). \\
src\_cnts\_aper\_b & 0.5-7 keV total net counts from the CSC catalog, with associated errors (src\_cnts\_aper\_b\_err). \\
src\_rate\_aper\_* & Count rates [cts s$^{-1}$] in various energy bands from the CSC catalog, with associated upper and lower limits (src\_rate\_aper\_lolim\_*, src\_rate\_aper\_hilim\_*). \\
HR\_CSC2p1 & Hardness ratio from CSC2.1, with lower and upper errors (HR\_CSC2p1\_errl, HR\_CSC2p1\_erru). \\
XMM\_CSC\_R & Cross-match radius between the associated XMM and Chandra sources. \\
S82X\_XID & Cross-match identifier with the Stripe 82X survey. \\
FFlux & 0.5-10 keV flux [erg s$^{-1}$ cm$^{-2}$] and associated error (FFlux\_Err). When a source is in more than one catalog, we follow the priority: 4XMMs, 4XMM, and CSC. \\
SFlux & 0.5-2 keV flux [erg s$^{-1}$ cm$^{-2}$] and associated error (SFlux\_Err). When a source is in more than one catalog, we follow the priority: 4XMMs, 4XMM, and CSC. \\
HFlux & 2-10 keV flux [erg s$^{-1}$ cm$^{-2}$] and associated error (HFlux\_Err). When a source is in more than one catalog, we follow the priority: 4XMMs, 4XMM, and CSC. \\
Lum\_Full & Derived 0.5-10 keV rest-frame luminosity, uncorrected for absorption [erg s$^{-1}$] and associated error (Lum\_Full\_err) for sources with spectroscopic redshift available. \\
Lum\_Soft & Derived 0.5-2 keV rest-frame luminosity, uncorrected for absorption  [erg s$^{-1}$] and associated error (Lum\_Soft\_err) for sources with spectroscopic redshift available. \\
Lum\_Hard & Derived 2-10 keV rest-frame luminosity, uncorrected for absorption  [erg s$^{-1}$] and associated error (Lum\_Hard\_err) for sources with spectroscopic redshift available. \\
PLUG\_RA, PLUG\_DEC & Right Ascension and Declination from the SDSS plug-plate catalog DR17. \\
z\_spec & Spectroscopic redshift of the source from SDSS DR17 or S82X. \\
Class & Spectral classification of the source from SDSS DR17 or S82X. \\
z\_origin & Origin of the redshift measurement and classification: ``S82 DR3" for values from \citealp{lamassa24}, ``CSC SDSS DR17" for values coming from the CSC association, and ``SDSS DR17" for those coming from our cross-matching. \\
NH & Logarithm of the absorbing column density derived using hardness ratios, with associated lower and upper bounds (NH\_lo and NH\_up). \\
\hline
\end{longtable}

\bibliography{sample631}{}
\bibliographystyle{aasjournal}

\end{document}